\documentclass[letterpaper,11pt]{article}
\usepackage{jheppub}
\pdfoutput=1
\usepackage{amsfonts}
\usepackage{amsmath}
\usepackage{amssymb}
\usepackage{comment}
\usepackage{float}
\usepackage{graphicx} 

\graphicspath{{./images/}}

\preprint{LITP-25-13}
\title{Phases of Supersymmetric Ground States in $AdS_4$}
\author{Harold Jones, Vineeth Krishna, Finn Larsen}
\affiliation{Leinweber Institute for Theoretical Physics, University of Michigan, Ann Arbor, MI, USA 48109}
\emailAdd{haljones@umich.edu}
\emailAdd{vkt@umich.edu}
\emailAdd{larsenf@umich.edu}

\abstract{
We construct the phase diagram of supersymmetric ground states in AdS$_4\times S^7$ supergravity. 
BPS black holes exist only when the conserved charges satisfy a certain non-linear constraint. For other charge sectors, we propose two component configurations comprised of a core black hole that carries macroscopic entropy, and a ``gas” that carries a macroscopic fraction of the charge. 

The superconformal index counts ground states only modulo a linear constraint on the charges. Each index line includes a pure BPS black hole but we find that, when differences between R-charge exceed a certain threshold, the index is dominated by a two component configuration. 

We illuminate this result by studying the gravitational path integral in Euclidean signature, as function of supersymmetric boundary conditions. We find that a pure BPS black hole saddle dominates the index when the R-charges are comparable, but not when they differ sufficiently. The region where the Euclidean path integral becomes unstable is precisely where two component configurations dominate the index. We show that the KSW criteria are insensitive to this instability.}

\begin{document}

\maketitle
\flushbottom

\section{Introduction}

In any physical system, it is important to determine the properties of the typical quantum states with minimal energy, given the conserved charges. In AdS$_4$ supergravity, the symmetry algebra permits unitary representations only when the energy satisfies a lower bound that is linear in the conserved charges: 
\begin{equation}
E\ell \geq  J + 2(Q_1\ell+Q_2\ell)~.
\label{eqn:introBPS}
\end{equation}
However, black holes can only saturate this bound if the charges satisfy the non-linear constraint 
\begin{equation}
    0 =  \left(Q_1 \ell + Q_2\ell\right)\left(\sqrt{1+\frac{64G^2}{\ell^4}(Q_1\ell)(Q_2\ell)} -1 \right)- J~. 
    \label{eqn:nlconstraint}
\end{equation}
This is surprising because, in a gravitational system, it is expected that typical states are black holes. After all, they have huge entropy, given by the area law, and they are organized by a few conserved charges, via Birkhoff’s theorem and its generalizations. 

There is a simple resolution to this tension: generically, a supersymmetric ground state has two components. One is a core black hole that carries large entropy. Another component, which we refer to as the ``gas”, carries the remaining macroscopic charge, but negligible entropy. 
Some aspects of this mechanism were recently developed in various dimensions, with and without supersymmetry \cite{Kim:2023sig,Choi:2024xnv,Bajaj:2024utv,Choi:2025lck,Larsen:2025jqo}. The purpose of this article is to study the two-component scenario in AdS$_4$ supergravity. 
One of our main results is Figure \ref{fig:cohom}, the phase diagram for the BPS states in AdS$_4$ that remain in the macroscopic limit. 

The basic two-component prescription identifies a classical ground state in any super-selection sector.
As in cases that were studied previously, there is a qualitative distinction between situations where the specified charges are ``overrotating” or ``overcharged”, relative to a pure BPS black hole. This corresponds to the right hand side of \eqref{eqn:nlconstraint} being positive (``mostly $Q$”) or negative (``mostly $J$"). The physical realization of the ``gas" is very different in these two cases. On the ``overrotating" side, which we call Region I, grey galaxies \cite{Kim:2023sig,Bajaj:2024utv,Choi:2025lck} carry angular momentum efficiently far from the black hole. 
On the ``overcharged" side, which we call Region II, the charge is carried by one or more dual giants \cite{Choi:2024xnv,Choi:2025lck}. This corresponds to a charged condensate. Nonetheless, the thermodynamic instability that underlies the two phases is very similar.

General black holes in gauged ${\cal N}=8$ supergravity have four distinct electric charges \cite{Hristov:2019mqp}, corresponding to the four Cartan generators of the $SO(8)$ isometry group of $S^7$. We study the ``two charge" case, where these four charges are equal in pairs. Accordingly, \eqref{eqn:introBPS} and \eqref{eqn:nlconstraint} are written in terms of two independent electric charges $Q_1$ and $Q_2$. This is significant, because the ``overcharged" region of the phase diagram supports two distinct phases, dependent on whether the two charges are comparable or very different. When the charge difference $Q_1-Q_2$ is small, the imbalance is carried entirely by the gas, so the core black hole has $Q_1=Q_2$. When it is larger, the gas carries only one charge, either $Q_1$ or $Q_2$. We call these regions of the phase diagram Region $II_a$ and $II_b$, respectively. 
In terms of symmetries, the two phases correspond to a gas component that carries $\frac{1}{4}$ or $\frac{1}{2}$ of maximal supersymmetry, respectively. 

The phase diagram of supersymmetric ground states has implications for the structure of the superconformal index \cite{WITTEN1982253,Kinney:2005ej}, the partition function where bosons and fermions are weighted with opposite signs. The index depends on two continuous variables, so it is codimension one in the space of charges. Each index line contains a unique point that corresponds to a BPS black hole, with charges that satisfy the constraint \eqref{eqn:nlconstraint}. For all other charge vectors on the same index line, the only candidate ground states are configurations with two component. Since black holes have large entropy, the simplest hypothesis identifies the extremum along an index line with the entropy of the pure BPS black hole that it intersects. Since our prescription assigns an entropy to all charge configurations, we can test this hypothesis. We find that the simple assumptions we make about a two-component supersymmetric ground state are inconsistent with a superconformal index that is dominated by a pure BPS black hole.   

Again, the balance between $Q_1$ and $Q_2$ is decisive. 
When the charges are comparable, the index is dominated by a pure BPS black hole. However, above a certain threshold, a two component configuration gives the largest contribution. 
Accordingly, we predict that the index exhibits a phase transition. Similar conclusions were previously reached for ${\cal N}=4$ SYM \cite{Choi:2025lck,Deddo:2025jrg}, but we consider its CFT$_3$ analogue. We present the phase diagram for the index in Figure \ref{fig:indphasediag}.

The Euclidean Gravitational Path Integral (EGPI) \cite{Gibbons:1976ue} offers a complementary perspective. The EGPI is a function of boundary conditions at the asymptotic boundary of spacetime that correspond to complex potentials conjugate to the conserved charges. Generally, there is one boundary condition for each conserved charge and the EGPI computes the grand canonical partition function. To identify the BPS sector, boundary conditions at infinity are restricted so that fermions are defined throughout spacetime, and then the EGPI determines the supersymmetric index \cite{Cabo-Bizet:2018ehj}. 

An implementation of this program considers the known classical black hole solutions and imposes supersymmetric boundary conditions, by selecting black holes with appropriate complex parameters. The resulting saddle points can be continued to Euclidean signature and then their on-shell action yields the indicial free energy as function of complex potentials. After Legendre transform to the microcanonical ensemble, the black hole entropy computed from the area law is recovered
\cite{Choi:2018fdc,Cabo-Bizet:2018ehj,Nian:2019pxj}.

This processing of the EGPI offers a path towards microscopic understanding of black hole entropy in AdS$_4$ spacetimes. After computing the supersymmetric index in the dual CFT$_3$ \cite{Choi:2019zpz}, the saddle point at large-$N$ yields an indicial free energy that is function of complex potentials that can be compared with the gravitational result. The successful comparison achieved this way is a major success of quantum gravity, as implemented by the AdS/CFT correspondence \cite{Maldacena:1997re,Witten:1998qj}. However, our results suggest that the agreement should be less general than the one that has been reported \cite{Choi:2019zpz}, because the index is dominated by a pure BPS black hole only in part of the phase diagram. To pursue this point, we consider several limitations of the EGPI. 

It is usually assumed that the gravitational path integral can be interpreted as a sum over quantum states in a Hilbert space, in analogy with quantum theory of systems without gravity. This property is far from obvious, and it is not even viable, unless the complex potentials that encode boundary conditions on the EGPI are consistent with convergence of the hypothesized sum over quantum states. This condition is nearly trivial, when the R-charges are equal: the only requirement is that the real part of the temperature is positive \cite{Chryssanthacopoulos}. However, the condition is more discerning when the charges are unequal. We find that it is satisfied if and only if the index is dominated by a pure BPS black hole. This result suggests that the supersymmmetric index computed from the CFT$_3$ must be sensitive to the phase transition as well.  

The EGPI is central to many recent research directions \cite{Saad:2019lba,Penington:2019kki,Almheiri:2019qdq,Iliesiu:2020qvm,Heydeman:2020hhw,Boruch:2022tno}, so it is important to understand its limitations without appeal to its interpreation as a sum over quantum states. Conceptually, the EGPI is a sum over ``all” geometries consistent with the boundary conditions. As such, it can be evaluated, at least approximately, by adding exponentials of the (negative) Euclidean action evaluated at all extrema, with appropriate determinants taking quantum corrections into account.  However, this prescription cannot be taken too seriously, it fails in simple examples \cite{Witten:2021nzp,Mahajan:2025bzo,Singhi:2025rfy}. In more complex circumstances, it is not clear {\it a priori} what saddle points should be taken into account.

KSW offers a principled prescription for the EGPI that is based on the underlying Lorentzian interpretion \cite{Kontsevich:2021dmb,Witten:2021nzp}. The asymptotic time, usually denoted $t$, is not appropriate for continuation $t\to -i\tau$, because there is an ergoregion, 
but a ``comoving” time $t-\Omega^{-1} \phi$ is viable, because it is globally defined outside the black hole horizon. This corresponds to positivity of thermodynamic factors such as $\beta(E - \Omega J)$, so it is related to the interpretation of the EGPI as a trace over quantum states. 
We find that, in our setting, the KSW condition is too weak: it is satisfied also in the unstable region of phase space. This complements other examples where it is too strong \cite{Maldacena:2019cbz,Bah:2022uyz,Chen:2023hra}, it excludes classical saddles that are thought to be legitimate. 

This article is organized as follows. 
In section \ref{sec:2chargesol}, we review the supersymmetric  black hole solutions in AdS$_4$ and their thermodynamics.  
In section \ref{sec:2comp} we discuss the principles governing two component solutions and construct the phase diagram of supersymmetric ground states.
In section \ref{sec:supindex} we use the results to determine the point on an index line that maximizes the entropy. 
In section \ref{sec:allowable} we study the Euclidean Gravitational Path Integral. We complexify the black hole solutions and study the resulting thermodynamics. 
We conclude in section \ref{sec:conclusion} with a brief summary, and discussion of open questions that are left for future research. 
In appendix \ref{sec:entext} we carry out the constrained extremization over the free energy of AdS$_4$ black holes.

\section{$AdS_4$ BPS Black Holes with Two Independent Charges}\label{sec:2chargesol}

In this section we introduce the Lorentzian AdS$_4$ black holes solutions we study. We specialize to the BPS limit. It does not exist unless the conserved charges satisfy the nonlinear constraint \eqref{eqn:nlconstraint}. This constraint defines a surface in the space of conserved charges that we call the black hole sheet.

\subsection{The Black Hole Solutions}

The vacuum of ${\cal N}=8$ gauged suergravity in four dimensions can be interpreted geometrically as AdS$_4\times S^7$.  
The $SO(8)$ isometry group of $S^7$ has rank $4$, corresponding to four commuting Killing vectors. From the 4D perspective this corresponds to an $SO(8)$ R-symmetry with a $U(1)^4$ Cartan subgroup. General black holes have independent electric charges under the four $U(1)$'s but, to keep calculations simple, we specialize to the case of pairwise equal charges.\footnote{This is the simplest case where one can expect a phase transition in the superconformal index, insofar as the analogy with the study of AdS$_5$ \cite{Choi:2025lck} holds up.} General black hole solutions in this superselection sector depend on the conserved charges $(M, J, Q_1, Q_2)$ where $M$ is the mass and $J$ is the angular momentum. 

The explicit black hole solutions \cite{Chong:2004na} depend on four parameters $(m,a,\delta_1,\delta_2)$ that have no obvious physical interpretation. They are related to the conserved charges through
\begin{align}
    M &= \frac{m(\cosh 2\delta_1 + \cosh 2\delta_2)}{2G (1-a^2 g^2)^2}~, \label{edef}\\
    J &= \frac{am(\cosh 2\delta_1 + \cosh 2\delta_2)}{2G (1-a^2 g^2)^2}~, \label{jdef}\\
    Q_1  &= \frac{m\sinh 2\delta_1}{4G(1-a^2 g^2)} ~, \label{q1def}\\
    Q_2 &= \frac{m\sinh 2\delta_2}{4G(1-a^2 g^2)} ~, \label{q2def}
\end{align}
where $G$ is Newton's gravitational constant in 4D. The gauge coupling $g$ of gauged supergravity is related to the $AdS_4$ radius $\ell$ as $g = \ell^{-1}$. The charges are normalized so that $2Q_i\ell$ are half integral for fermions and integral for bosons\footnote{The explicit factor of $2$ is due to the relation $\ell_{S^7} = 2\ell$. The $Q_i$ are angular quantum numbers on $S^7$ and so they are half integer quantized in units of the $S^7$ radius $l_{S^7}$.}. 

The explicit metric is\footnote{
The notations $\rho^2_{\rm here}=W_{\rm there}$ and
$\phi_{\rm here} = (1-a^2g^2) \phi_{\rm there}$ recovers the metric in \cite{Chong:2004na} and simplifies the reduction to the special case $\delta_1=\delta_2$ discussed in literature \cite{Kostelecky:1995ei,Caldarelli:1999xj,Papadimitriou:2005ii,Larsen:2020lhg}.   
``The" parameter $m$ that is common in the single charge context is related to the one here through: $m_{there} = m_{here} (1 + s^2_1 + s^2_2)$. Our conventions agree with \cite{Cassani:2019mms}.}:
\begin{eqnarray}
ds^2   & =&  - \frac{\Delta_{r^\prime}}{\rho^2}\left(dt - \frac{a\sin^2\theta}{1-a^2g^2} d\phi\right)^2 + \frac{\Delta_\theta\sin^2\theta}{\rho^2} \left(a~ dt - \frac{r_1 r_2 + a^2}{1-a^2 g^2} d\phi\right) ^2\cr
&&+ \frac{\rho^2}{\Delta_{r^\prime}}dr^\prime {}^2 +  \frac{\rho^2}{\Delta_\theta}d\theta^2~,
\label{eqn:Lormetric}
\end{eqnarray}
where 
\begin{eqnarray}
\Delta_\theta & = & 1 - a^2g^2 \cos^2\theta~,
\label{eqn:deltather}\\
\rho^2 & = & r_1 r_2 + a^2 \cos^2\theta~, 
\label{eqn:rhodef}\\
\Delta_{r^\prime} & = & r^\prime{}^2 + a^2- 2mr^\prime + g^2 r_1 r_2 (r_1 r_2 + a^2)~,
\label{eqn:deltar}
\end{eqnarray}
and
\begin{equation}
r_i = r^\prime + 2 \,m\, s^2_i~.
\label{eqn:shifteddef}
\end{equation}
Going forward we will use a shifted radial coordinate defined by
\begin{equation}\label{rdef}
    r = r^\prime +m(s_1^2+s_2^2)~.
\end{equation}
The quantities $r_{1,2}$ in terms of the new radial coordinate become
\begin{equation}
    r_{1,2} = r\pm m(s_1^2-s_2^2)~.
\end{equation}
We use the abbreviations $s_1 = \sinh\delta_1,~c_1=\cosh\delta_1$ and  $s_2 = \sinh\delta_2, ~c_2=\cosh \delta_2$ frequently here and in the following. 
The geometries are supported by the gauge fields
\begin{equation}
A_1 = \frac{2ms_1 c_1 r_2}{\rho^2}\left( dt  - \frac{a\sin^2\theta}{1-a^2g^2}d\phi \right)~, \qquad A_2 = \frac{2ms_2 c_2 r_1}{\rho^2}\left( dt  - \frac{a\sin^2\theta}{1-a^2g^2}d\phi \right)~,
\label{eqn:Lorgauge}
\end{equation}
and the scalar fields 
\begin{align}
e^\varphi & = 1+\frac{r_1(r_1-r_2)}{\rho^2}~,\cr
\chi & = \frac{a(r_2-r_1)\cos\theta}{r^2_1+a^2\cos^2\theta}~.
\end{align} 

The black hole entropy computed from the area law is given by
\begin{equation}
    S = \frac{\pi}{G} \frac{r_{1+}r_{2+}+a^2}{1-a^2g^2}~,
    \label{eqn:BHentropy}
\end{equation}
and the thermodynamic potentials conjugate to the conserved charges are: 
\begin{align}
\beta & = \frac{4\pi (r_{1+} r_{2+} +a^2)}{\partial_r \Delta_r\Big|_{r_+}}
\label{eqn:betapar} \\
\Omega & = \frac{a(1+ g^2 r_{1+} r_{2+})}{r_{1+} r_{2+} + a^2}~,
\label{eqn:omegapar} 
\\
\Phi_1 & = \frac{2m s_1 c_1 r_{2+}}{r_{1+} r_{2+} + a^2}~,
\label{eqn:phi1par} 
\\
\Phi_2 & = \frac{2m s_2 c_2 r_{1+}}{r_{1+} r_{2+} + a^2}~.
\label{eqn:phi2par} 
\end{align}
In all these formulae $r_+$ is the coordinate position of the event horizon, given by the largest positive root of the quartic equation $\Delta_r(r)=0$ and $r_{1+},r_{2+}$ are the values of $r_{1,2}$ evaluated at $r=r_+$. The definition of $r_+$ through $\Delta_r(r_+)=0$ relates $m$ and $r_+$. Thus, the solution and its physical properties depend on four continuous parameters $(a, \delta_1, \delta_2, m)$ or equivalently $(a,\delta_1,\delta_2,r_+)$.

\subsection{The BPS Limit}
Supersymmetry of the Lorentzian theory 
imposes the BPS bound on the four conserved charges: 
\begin{equation}
M\ell - J - 2(Q_1\ell + Q_2\ell) \geq 0~. 
\label{eqn:BPSineq}
\end{equation}
The parametric formulae 
(\ref{edef}--\ref{q2def}) 
give: 
\begin{equation}
M\ell - J - 2(Q_1\ell + Q_2\ell) =  \frac{m\ell}{G(1-a^2g^2)} \Big( \frac{1 + s^2_1 + s^2_2}{1+ag}  - (s_1 c_1 + s_2 c_2)\Big)
~.
\end{equation}
Let us define the ratio
\begin{equation}
    \delta_{12} \equiv  \frac{1+s^2_1 + s^2_2}{s_1 c_1 + s_2 c_2} 
=\frac{s_1 c_1 - s_2 c_2}{s^2_1 - s^2_2} = \coth{(\delta_1+\delta_2)}~.\label{eqn:ratio}
\end{equation}
The three different forms of the right hand side \eqref{eqn:ratio} are related by standard hypergeometric identities. 
Therefore, the BPS bound is saturated when $a$ is given through: 
\begin{equation}
1 + ag  = \delta_{12}~.
\label{eqn:BPScon}
\end{equation}

The relation \eqref{eqn:BPScon} between the four parameters $(m, a, \delta_1, \delta_2)$ is necessary for supersymmetry, but
it is insufficient. To show this, we
introduce the variables $(q,r_*,\delta_{12})$ and define them in terms of the existing parameters through 
\begin{align}
q &= m(s_1 c_1 + s_2 c_2)= m \sinh(\delta_1 + \delta_2)~, 
\label{eqn:qdef}\\
r^2_* &= ag^{-1} + m^2(s^2_1-s^2_2)^2~,
\label{eqn:rstardef}\\
\delta_{12} &= \coth{(\delta_1+\delta_2)}~.
\end{align}
The four parameters of the solutions now become $(a,r_*,q,\delta_{12})$. We rewrite the conformal factor $\Delta_r$ \eqref{eqn:deltar} in terms of the new parameters as\footnote{Note that this expression is written in terms of the shifted radial coordinate $r$ defined in \eqref{rdef}} 
\begin{eqnarray}
 \Delta_r (r) & = & \Big(q - r \,\delta_{12}\Big)^2 + g^2(r^2 - r^2_*)^2 + \left((1+ag)^2-\delta_{12}^2\right)(r^2-r_*^2+a g^{-1}) 
~,
\label{eqn:Delr}
\end{eqnarray}
which, for the BPS solutions satisfying \eqref{eqn:BPScon} can be written as a sum of two squares
\begin{equation}
    \Delta_r^* (r)  = \Big(q - r \,\delta_{12}\Big)^2 + g^2(r^2 - r^2_*)^2~.
\label{eqn:DelrBPS}
\end{equation}
To solve the horizon equation $\Delta_r^*(r_+)=0$ for the three parameter black holes satisfying \eqref{eqn:BPScon}, we need to impose
\begin{align}
    q &= r_+ \, \delta_{12}~,\\
    r_+ &= r_*~,
\end{align}
if all the parameters are taken to be real. The first equation relates the parameter $q$ to $r_+$. The second equation imposes additional constraint which eliminates the variable $r_*$. The combination of this new condition and the formula \eqref{eqn:BPScon} for $a$ guarantees $q=(1+ag)r_+$. Therefore, the real BPS black holes in Lorentzian signature form a co-dimension two-surface in the parameter space $(m, a, \delta_1, \delta_2)$. These solutions were first written down in \cite{Cvetic:2005zi}, which can also be thought of as taking the appropriate supersymmetric limit of the solutions of \cite{Chong:2004na}.

We can interpret the two parameter family of BPS black holes as a surface in the three dimensional space of all BPS configurations that is defined by the conserved charges $(J, Q_1, Q_2)$. In this context we refer to the BPS black holes as the black hole {\it sheet}. In Section \ref{sec:allowable} we will interpret the two parameter family of real BPS black holes as a surface in a space involving complex variables that we will refer to as the {\it real} black hole surface. 

We can label the BPS black holes by the two variables $(\delta_1,\delta_2)$. To make this explicit, recall that 
BPS black holes satisfy $q=(1+ag)r_+$. Inserting \eqref{eqn:BPScon}, \eqref{eqn:qdef}, and \eqref{eqn:rstardef} for 
 $a$, $q$, and $r_+=r_*$, we find an equation for $m$ that gives
\begin{equation}
    (mg)^2 = \frac{\cosh^2(\delta_1+\delta_2)}{e^{\delta_1+\delta_2}\sinh^3(\delta_1+\delta_2)\sinh(2\delta_1)\sinh(2\delta_2)}~. 
    \label{mbps}
\end{equation}
Then \eqref{eqn:BPScon} and \eqref{eqn:rstardef} easily give
\begin{equation}
   g^2r_*^2 =  g^2r_+^2 = \frac{\tanh\delta_1\tanh\delta_2}{e^{\delta_1+\delta_2}\sinh^3(\delta_1+\delta_2)}~. \label{rpbps}
\end{equation}

Inserting \eqref{eqn:BPScon} and \eqref{mbps} for $a$ and $m$ in the formulae (\ref{jdef} 
-\ref{q2def}), we can express the conserved charges $(J, Q_1, Q_2)$ in terms of the two parameters $(\delta_1, \delta_2)$. This gives an explicit formula for the constraint between the conserved charges that is satisfied by all real BPS black holes: 
\begin{equation}
    J = J_c(Q_1,Q_2) = \left(Q_1 \ell + Q_2\ell\right)\left(\sqrt{1+\frac{64G^2}{\ell^4}(Q_1\ell)(Q_2\ell)} -1 \right)~. 
    \label{eqn:Jconstraint}
\end{equation}
On the black holes sheet, the black hole entropy \eqref{eqn:BHentropy} similarly becomes 
\begin{equation}
    S = \frac{\pi \ell^2}{2G} \frac{J}{Q_1\ell+Q_2\ell}~.
\end{equation}
The expression on the right hand side is defined only modulo \eqref{eqn:Jconstraint}.

\begin{figure}[t]
    \centering
    \includegraphics[width=0.6\linewidth]{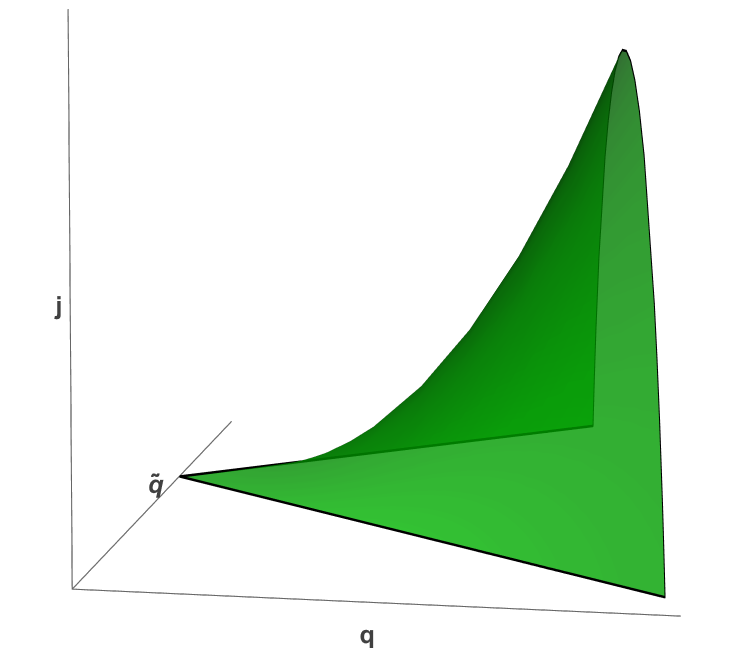}
    \caption{A plot of the ``black hole sheet''. It has boundaries at the bottom of the plot where it intersects the $j=0$  plane at the two lines $q=|\tilde q|$. As $j$ increases, the surface shrinks inwards from these lines. The lighter shade of green is in the foreground. The darker shade is where the sheet has bent over and there is a second branch hidden behind the translucent foreground. See Figure \ref{fig:constj} for a constant $j$ slice of the black hole sheet.}
    \label{fig:BHsheet}
\end{figure}

\subsection{The Black Hole Sheet}\label{sec:bhsheet}
The AdS/CFT correspondence relates the gravitational constant $G$ and the microscopic parameter $N$ through the dimensionless ratio:
\begin{equation}
\label{eqn:kdef}
    k = \frac{\ell^2}{2G} = \frac{\sqrt 2}{3} N^\frac32~.
\end{equation}
The $k$ sets the scale for all macroscopic variables. It is convenient to introduce dimensionless \textit{densities} of the electric charges, angular momentum, and entropy 
\begin{align}
\label{q1q1def}
    q_{1,2} &= k^{-1}(Q_{1,2}\,\ell) ~,\\
  j &= k^{-1}J ~,\\
    s &= k^{-1} S~. 
\end{align}
It will also be natural to use the linear combinations of electric charges that give the total charge and the asymmetry between the two charges: 
\begin{align}
    q &= q_1+q_2~,\label{linpl}\\
    \tilde q &= q_1-q_2~.\label{linmin}
\end{align}

Rewriting the constraint \eqref{eqn:Jconstraint} in terms of 
charge densities, we find: 
\begin{align}\label{sheetdef}
    j = j_c(q,\tilde q) = q\left(\sqrt{1+4\left(q^2-\tilde q^{2}\right)}-1\right)~.
\end{align}
This defines the 2-dimensional ``black hole sheet" in the 3D charge space parameterized by ($q,\tilde q,j$). BPS black holes exist only on the black hole sheet. Figure \ref{fig:BHsheet} plots the black hole sheet as a surface in the 3D charge space. 

Let us take a closer look at this sheet. We note that the black hole sheet has boundaries. It \textit{ends} on the lines $q=|\tilde {q}|$, which lie on the $j=0$ plane. The black hole sheet looks like a corner of an infinite tent which is pinned down to the $j=0$ plane along the lines $q=\pm \tilde q$.

\begin{figure}[t]
    \centering
    \includegraphics[width=0.6\linewidth]{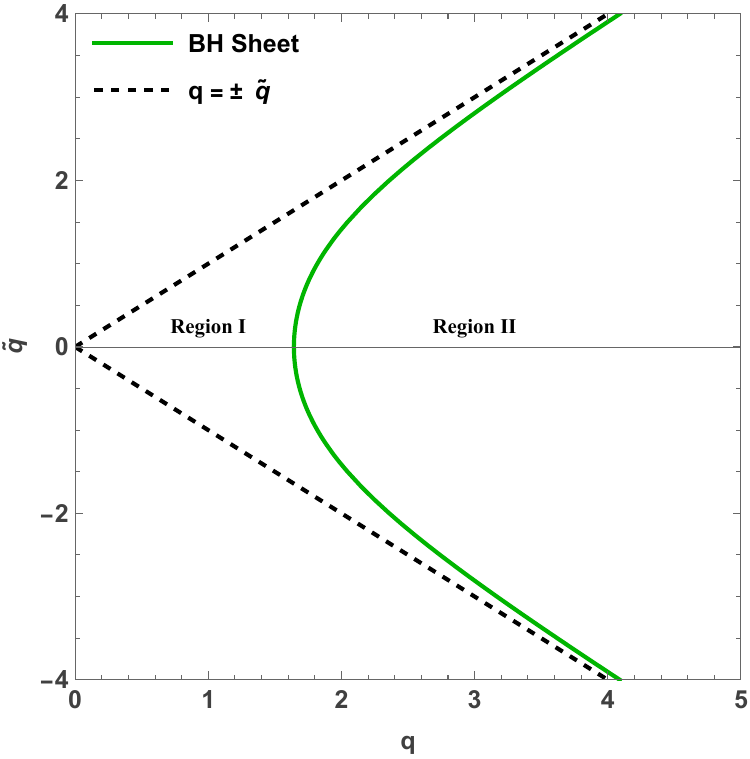}
    \caption{Constant $j$ cross section of the black hole sheet, represented by the green curve. The curve moves to the right for larger $j$. Conversely, it moves left for smaller $j$, approaching the dashed lines as $j\to 0$. The surface formed by the dashed lines at all values of $j$ form the boundary of the allowed region. The black hole sheet splits this region into two components -- region $I$ and region $II$. The norm of the slope of the green curve is $>1$ at every point and asymptotically approaches unity at large charges.  }
    \label{fig:constj}
\end{figure}

It is also instructive to examine slices of charge space with constant $j$, as illustrated in Figure \ref{fig:constj}. The green line is the intersection of the slice with the sheet. The cross-section of the sheet always has slope $>1$ and asymptotes to the $q=\pm \tilde q$ lines at infinity. On slices with larger $j$, the green curve recedes to the right. Conversely, on slices with smaller $j$, it moves to the left. In the limit $j\to 0$ the green curve approaches the asymptotes $q=|\tilde q|$. The representation of the black hole sheet by constant $j$ slices will prove valuable in the next section when we present results of calculations.

The entropy density of the black holes can be presented as the simple formula
\begin{equation}\label{bhentropy}
    s_{BH}(j,q) = \pi \frac{j}{q}~.
\end{equation}
However, one can equally well use the charge constraint \eqref{sheetdef} to convert the expression \eqref{bhentropy} to a function of any two variables formed from $(q,\tilde q,j)$. For example, the entropy density of the black hole in terms of the charges $(q,\tilde q)$ is given by,
\begin{equation}\label{bhentropyqqt}
    s_{BH}(q,\tilde q) = \pi \left(\sqrt{1+4(q^2-\tilde q^2)}-1\right) ~.
\end{equation}
At fixed $j$, \eqref{bhentropy} shows the entropy of the black hole is maximized when $q$ takes its minimum value. From the Figure \ref{fig:constj} it is clear that this happens at the tip of the green curve where it intersects the $\tilde q = 0$ axis. The $q$ coordinate of this point can be computed by substituting $\tilde q = 0$ in \eqref{sheetdef} and solving for $q$. Since the equation is quartic in $q$ the resulting expression is complicated, we will not be reporting it here.

\section{Two Component Supersymmetric Phases in AdS$_4$}\label{sec:2comp}
In this section, we construct the microcanonical phase diagram of supersymmetric, electrically charged, rotating black holes in supergravity on $AdS_4 \times S^7$. At charges where there are no known supersymmetric solutions, we identify new phases with two components --- a ``core'' black hole and a non-interacting ``gas'' that exists in the black hole background. 

\subsection{The Supersymmetric Phase Diagram: Principles}\label{sec:princi}
A supersymmetric solution to 11D supergravity with two pairs of equal charges saturates the BPS bound \eqref{eqn:BPSineq}, so it satisfies a linear relation between the mass, angular momentum and the electric charges of the solution
\begin{equation}\label{bpsbound}
    M \ell = J + 2(Q_1\ell + Q_2\ell)~.
\end{equation}
In writing this mass formula we have chosen a specific supercharge which is preserved in the solutions. It is $\mathcal{Q}_{-}^{++++}$ with the quantum numbers $$(M\ell,J,2Q_1\ell,2Q_2\ell,2Q_3\ell,2Q_4\ell)=\left(\frac12,-\frac12,\frac12,\frac12,\frac12,\frac12\right)~.
$$ 
The BPS solutions that preserve this supercharge have positive conserved charges in the sense that
\begin{equation}\label{allowed}
    J\geq 0 ~,~~~ Q_1\ell\geq 0 ~,~~~ Q_2\ell \geq 0~. 
\end{equation}
This can also be written as
\begin{equation}\label{allowed2}
    J\geq 0 ~,~~~ Q\ell \geq |\tilde Q\ell|~.
\end{equation}
We call the region defined by \eqref{allowed} or \eqref{allowed2} the \textit{allowed region}. 

The conserved charges of the pure BPS black hole solutions described in section \ref{sec:2chargesol} satisfy the non-linear constraint \eqref{eqn:Jconstraint}, as well the linear mass relation \eqref{bpsbound}. Therefore, they exist only on a co-dimension $1$ surface that we call the black hole sheet. This surface splits the allowed region \eqref{allowed2} into two disconnected components -- Region $I$ and Region $II$. 
These regions, separated from each other by the black hole sheet, are illustrated in Figure \ref{fig:constj}. 

In component $I$ of the allowed region, the angular momentum of any point is larger than that of the corresponding point on the black hole sheet with the same electric charges, so it can be characterized as the set of points for which the charge densities of the point $(q,\tilde q, j)$ satisfy
\begin{equation}\label{reg1def}
    \text{Region $I$ : }~~~~~~~~~j>j_c(q,\tilde q), ~~~~q > |\tilde q| ~,
\end{equation}
where $j_c(q,\tilde q)$ is the non-linear function defined in \eqref{sheetdef}. Region $I$ lies ``above'' and to the left of the black hole sheet in Figure \ref{fig:BHsheet}. It is bounded by the planes $q=\pm \tilde q$, in addition to the black hole sheet. 

In component $II$ of the allowed region, the angular momentum of any point in the region is less than that of the corresponding point on the black hole sheet at the same values of the electric charges. In other words, region $II$ is defined as the set of points which satisfy
\begin{equation}\label{reg2def}
    \text{Region $II$ : } ~~~~~~~~~ 0<j<j_c(q,\tilde q)~.
\end{equation}
Equivalently, region $II$ is where the electric charge density $q$ is greater than that of the corresponding point on the black hole sheet at the same value of $(j,\tilde q)$. This follows because $j$ and $q$ are monotonically related on the black hole sheet $j=j_c(q,\tilde q)$, when $\tilde{q}$ is kept fixed. It also follows from inspection of Figures \ref{fig:BHsheet} and \ref{fig:constj}.

The preceding geometric description of the three dimensional charge space anticipates a supersymmetric phase diagram where 
Regions $I$ and $II$ correspond to distinct phases. 
However, the only explicitly known supersymmetric configurations with $O(1/G)$ entropy are the pure BPS black hole solutions discussed in Section \ref{sec:2chargesol}, and they exist only on the black hole sheet \eqref{sheetdef}. 
In the rest of the section we discuss new supersymmetric configurations that also carry $O(1/G)$ entropy, and cover Regions $I$ and $II$ in their entirety. These new supersymmetric configurations are constructed as combinations of two distinct supersymmetric components -- a ``core'' BPS black hole and a BPS ``gas'' -- that co-exist. The spacetime realization of such supersymmetric composites, and even their existence, is far from obvious, but we defer the discussion of these points to later in this section. 

For now, we will simply state some basic properties that we assume can be satisfied by some two component configurations. That will be sufficient to construct a supersymmetric phase diagram in the three dimensional charge space spanned by $(q,\tilde q, j)$. The properties of the two component configurations we assume are the following: 
\begin{itemize}
    \item[(i)] The two components are non-interacting: the back reaction of the gas on the black hole geometry, and vice versa,  is negligible. Therefore, the total conserved charges of configurations composed of a core ``black hole'' and a surrounding ``gas'' is the simple sum of the individual contributions. 
    The BPS relation \eqref{bpsbound} is linear in the conserved quantities so, if it is satisfied by the individual components, it is also satisfied by the combined system. 
    \item[(ii)] The entropy of the gas component is negligible. More precisely, the gas entropy is subleading in units of the gravitational constant $G$, when compared to the $O(1/G)$ classical entropy of the black hole.
    \item[(iii)] The gas component acts as a sink for angular momentum or electric charge. There is no upper bound on the amount of charge or angular momentum that the gas component can carry.
    \item[(iv)] The charge densities of the gas component by itself satisfy the lower bounds 
    \begin{equation}
        q_{1g}\geq 0 ~,~~~ q_{2g} \geq0 ~,~~~ j_g \geq 0~, 
    \end{equation}
    which, in terms of the linear combinations \eqref{linpl} and \eqref{linmin}, is equivalent to
    \begin{equation}\label{gasconst}
        q_g \geq |\tilde q_g| ~,~~~ j_g \geq 0~.
    \end{equation}
    The inequalities above define a three sided cone in the charge space whose boundaries are given by the three planes $q_{g}=\pm \tilde   q_g$ and $j_g=0$.
    \item[(v)] The core black hole must be one of the pure BPS black hole solutions whose charges lie on the black hole sheet \eqref{sheetdef}.
\end{itemize}
In the following subsections we apply these assumptions and construct the supersymmetric phase diagram in the microcanonical ensemble. We achieve this by identifying the configuration with the highest entropy from the set of all possible two component solutions which carry the given set of total charges, subject to the constraints \eqref{gasconst}. 

\subsection{Stability of Pure BPS Black Holes}\label{sec:reg0}

As a step towards the phase diagram we first show that the BPS black holes described in section \ref{sec:2chargesol} are stable against the formation of any two component configuration that satisfies the assumptions listed near the end of subsection \ref{sec:princi}.

Consider a point $P$ on the black hole sheet with charge densities $(q,\tilde q, j)$. The entropy of the pure BPS black hole at $P$ \eqref{bhentropy} can be written as
\begin{equation}\label{reg0ent}
    s(q,\tilde q) = \pi \left(\sqrt{1+4(q^2-\tilde q^2)}-1\right) ~,
\end{equation}
after eliminating $j$ using the nonlinear constraint \eqref{sheetdef}. A two component configuration at $P$ carries the same total charges $(q,\tilde q, j)$, but they are shared between the core black hole and the gas. Assuming temporarily that the gas component has infinitesimal charges $(q_g,\tilde q_g) =(-\delta q,-\delta \tilde q)$, the charges of the core black hole differ from those of the pure BPS black hole by $(\delta q, \tilde \delta q)$, In our conventions $\delta q$ is negative, since the gas component carries positive $q_g$. 
Positivity of the gas charges further imposes the inequality \eqref{gasconst}, so changes to the core black hole charges satisfy: 
\begin{equation}\label{reg0gasin}
    |\delta q| > |\delta \tilde q|~.
\end{equation}

According to our assumptions, the total entropy of the two component system is entirely due to the core black hole, so it differs from that of the pure BPS black hole by
\begin{equation}
    \delta s = \left.\frac{\partial s}{\partial q}\right|_{P} \delta q + \left.\frac{\partial s}{\partial \tilde q}\right|_{P} \delta \tilde q = \frac{\pi}{\sqrt{1+4(q^2-\tilde q^2)}} (q \delta q 
    -\tilde q\delta\tilde q)~. 
\label{entvar} 
\end{equation}
The sign of $\delta s$ is determined by the sign of the quantity in the brackets in \eqref{entvar}. Since $P$ lies on the black hole sheet, which is completely inside the allowed region defined by \eqref{allowed2}, the coordinates at $P$  satisfy
\begin{equation}\label{reg0bhin}
    q > |\tilde q|~.
\end{equation} 
Combining \eqref{reg0gasin} and \eqref{reg0bhin}, we find that the quantity in the brackets of \eqref{entvar} is negative, and so
\begin{equation}\label{concl}
    \delta s < 0~.
\end{equation}
Therefore, the pure BPS black hole at point $P$ on the black hole sheet cannot increase its entropy by emitting an infinitesimal amount of gas in any allowed direction \eqref{reg0gasin} of the charge space. 

\begin{figure}
    \centering
    \includegraphics[width=0.6\linewidth]{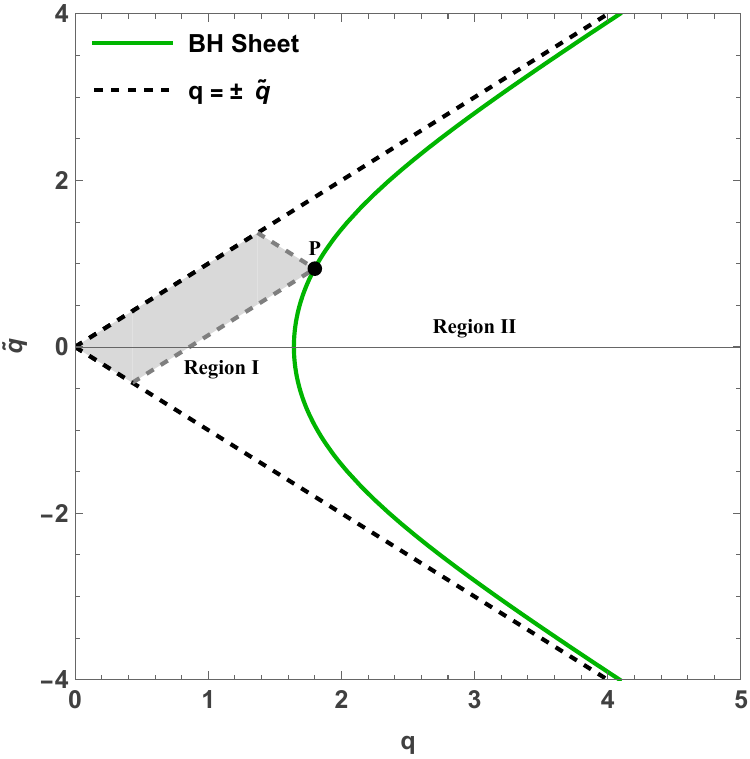}
    \caption{A projection of the charge space to a slice with constant $j$. The black hole sheet $j = j_c(q,\tilde q)$ given in \eqref{sheetdef} is defined inside the wedge created by the two black dashed lines $q=\pm \tilde q$. The green curve is the cross section of the black hole sheet at the same value of $j$ as the point $P$. The part of the black hole sheet with smaller values of $j$ lies to the left of the green curve, and the part with larger values is to the right. The grey shaded region is on the black hole sheet and contains all the core BPS black holes that can form a two component configuration with total charges at point $P$.}
    \label{fig:2compposssheet}
\end{figure}

This conclusion generalizes to two component configurations where the gas carries a finite fraction of the charges. Suppose the total charges $(q,\tilde q, j)$ at the point $P$ are shared between a core BPS black hole with charges $(q_c,\tilde q_c,j_c)$ and a gas component. Positivity of the gas charges restrict the black hole charges to a region on the black hole sheet that satisfies
\begin{align}
    q - q_c&\geq |\tilde q - \tilde q_c|~, \label{reg0def1}\\
    j - j_c &\geq 0~. \label{reg0def2}
\end{align}
This region is depicted as the grey shaded patch in Figure \ref{fig:2compposssheet}. This patch of the black hole sheet 
is entirely at smaller $j$ than the green curve which contains $P$. Therefore, whenever the gas component has positive charges, in the sense of \eqref{reg0def1}, it also has positive angular momentum, and so \eqref{reg0def2} is satisfied. We need to compare the entropy of a pure BPS black hole at $P$ and a two component configuration with core black hole somewhere in the shaded region. A curve relating the two candidates can be chosen so $(q,\tilde{q})$ vary monotonically. The entropy decreases for infinitesimal motion along such a curve, so it decreases for the full displacement. 
We conclude that the pure BPS black hole at the point $P$ has larger entropy than any allowed two component configuration
with the same total charges.


\subsection{The Supersymmetric Phase: Region $I$}\label{sec:reg1}
\begin{figure}
    \centering
    \includegraphics[width=0.6\linewidth]{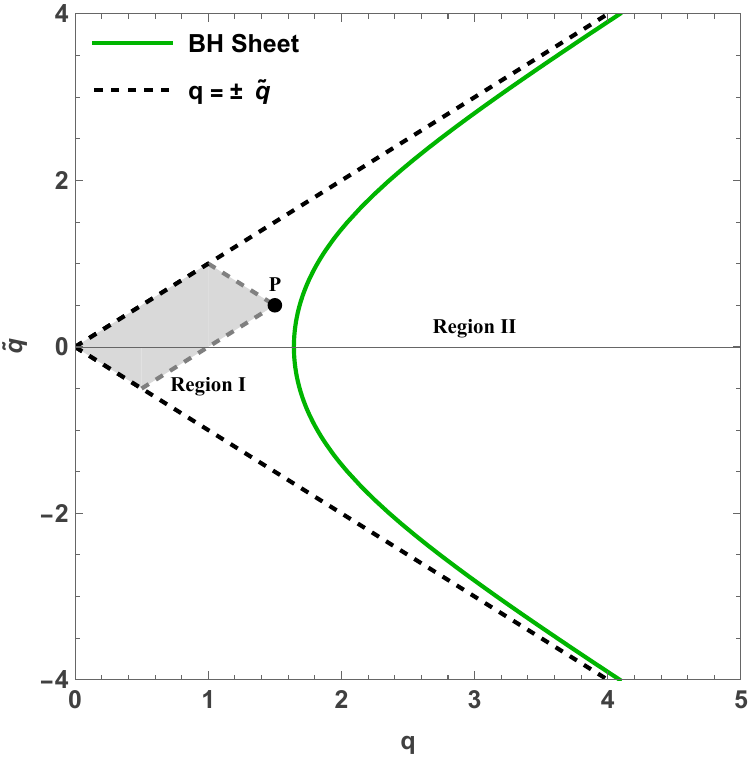}
    \caption{A projection of the charge space to a slice with constant $j=j_P$ that includes the point $P$ in region $I$. The green curve is the cross section of the black hole sheet at $j=j_P$. The grey shaded region is on the black hole sheet and contains all the core black holes that can form a two component configuration at point $P$. This region of the black hole sheet is bounded by the boundaries of the allowed region (black dashed lines) and the boundaries of the gas constraints (grey dashed lines). Since the grey shaded region lies to the left of the green curve, it only contains points whose $j$ value is less than that of the point $P$. The point on the black hole sheet that is directly below the point $P$ has the maximum entropy in the shaded region.}
    \label{fig:2compposs}
\end{figure}

Region $I$ is on the large angular momentum side of the black hole sheet and defined precisely by \eqref{reg1def}. Consider a point with charge densities $(q,\tilde q, j)$ in the interior of Region $I$, denoted by $P$ in Figure \ref{fig:2compposs}. There are no pure black holes with these charges, since $P$ is not on the black hole sheet. The candidate two component configurations with the total charges $P$ consists of a core BPS black hole with the charge densities $(q_c,\tilde q_c, j_c)$ on the black hole sheet and a gas component that carries the rest of the charges. The conditions that the charges of the gas are positive \eqref{gasconst} imposes 
\begin{align}
    q - q_c&\geq |\tilde q - \tilde q_c|~, \label{reg1def1}\\
    j - j_c &\geq 0~, \label{reg1def2}
\end{align}
on the core black hole. The second of these is automatically satisfied on the large angular momentum side of the black hole sheet. The points $(q_c,\tilde q_c,j_c)$ that satisfy the bounds \eqref{reg1def1} and \eqref{reg1def2} are indicated as a shaded region on Figure \ref{fig:2compposs}. 

According to our assumptions, the entropy of a two component configuration is equal to the entropy \eqref{bhentropy} of the core black hole. The dominant phase at the point $P$ in region I has a core black hole whose entropy \eqref{reg0ent} is maximized over the domain defined by \eqref{reg1def1}. Consider the candidate two component configuration at point $P$ whose core black hole carries the same electric charges $(q,\tilde q)$ as point $P$ i.e. the gas component carries only angular momentum. The arguments in Section \ref{sec:reg0} showed that such a core black hole cannot increase its entropy shedding electric charge and moving into the interior of the patch defined by \eqref{reg1def1}. Therefore, the maximum entropy configuration at any point in region $I$ is the one in which the gas component carries only angular momentum. This result populates every point in Region $I$ with a two component phase whose gas component carries only angular momentum.

\subsection{The Supersymmetric Phase: Region $II$}\label{sec:reg2}
Region $II$ is on the large charge side of the black hole sheet and defined precisely by \eqref{reg2def}. It contains all points which are to the right side and below the black hole sheet in Figures \ref{fig:BHsheet} and \ref{fig:constj}. 

\begin{figure}
    \centering
    \includegraphics[width=0.6\linewidth]{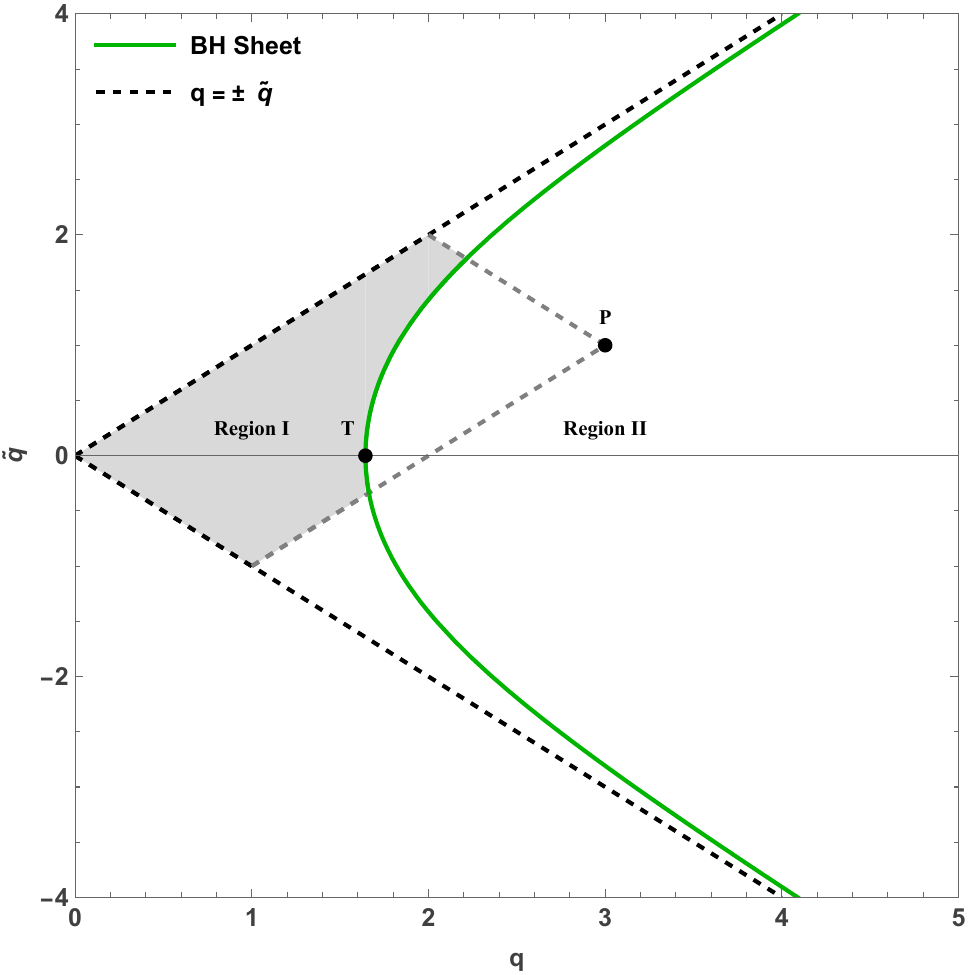}
    \caption{A projection of the charge space to a constant $j$ slice with a point $P$ in region $IIa$. The grey shaded surface is on the black hole sheet and contains the core black holes that can combine with a gas to form a two component configuration with total charges $P$. This region of the black hole sheet is bounded by the boundaries of the allowed region (black dashed lines), the boundaries of the gas constraints (grey dashed lines) and the black hole sheet (green curve). The two component configuration at point $P$ with the maximum entropy contains a core black hole at Point $T$.}
    \label{fig:2comppossreg2}
\end{figure}

\begin{figure}
    \centering
    \includegraphics[width=0.6\linewidth]{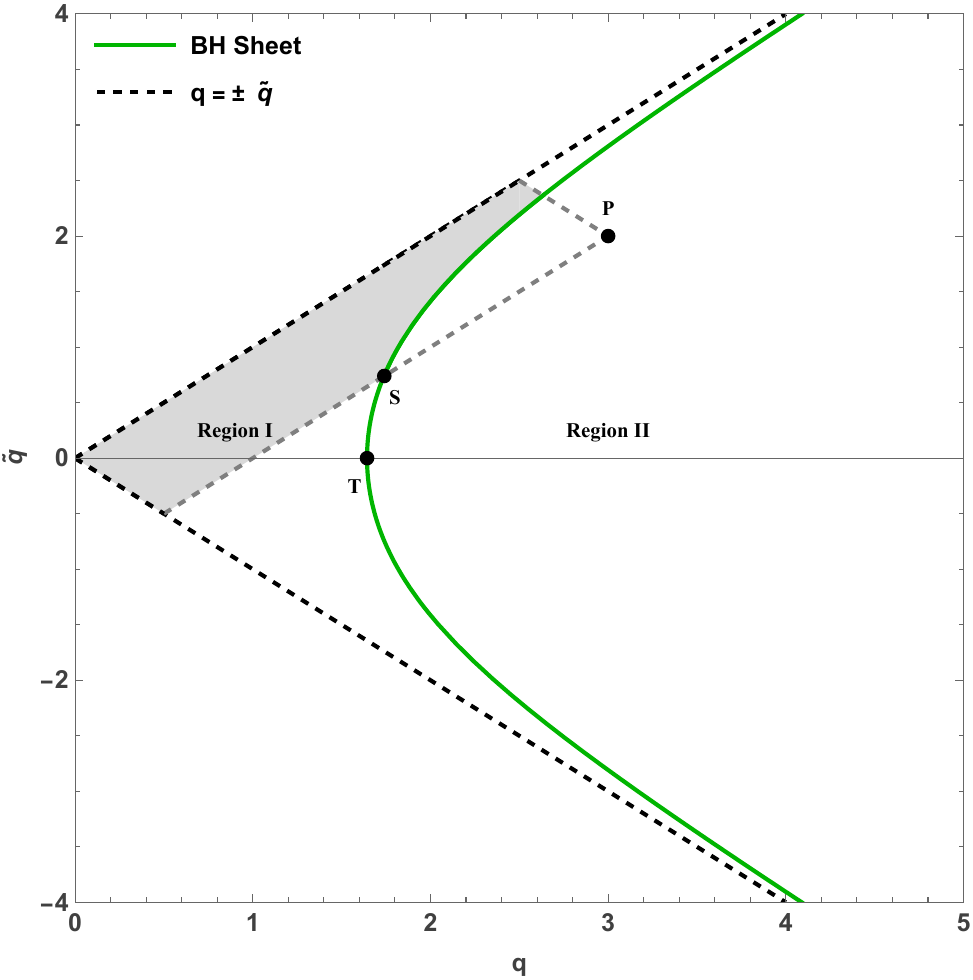}
    \caption{A projection of the charge space to a constant $j$ slice with a point $P$ in region $II_b$. The grey shaded surface contains all the core black holes on the sheet that can combine with a gas to form a two component configuration at point $P$. This region of the black hole sheet is bounded by the boundaries of the allowed region (black dashed lines), the boundaries of the gas constraints (grey dashed lines) and the black hole sheet (green curve). The two component configuration at point $P$ with the maximum entropy contains a core black hole at Point $S$, since point $T$ is not accessible.}
    \label{fig:2comppossreg2b}
\end{figure}

Consider a point $P$ in region $II$ with charge densities $(q,\tilde q, j)$. The two component configurations with this total charge have a core black hole with charge densities $(q_c,\tilde q_c, j_c)$ that are on the black hole sheet. Positivity of the gas charges again impose \eqref{gasconst} which corresponds to the bounds (\ref{reg0def1}-\ref{reg0def2}) on the charges of the core black hole. In region $II$ the condition that the gas angular momentum is non-negative can be satisfied only if the gas carries electric charge. The allowed core black holes are inside a cone in the space of electric charges. This is the grey shaded region in Figures \ref{fig:2comppossreg2} and \ref{fig:2comppossreg2b} which correspond to two distinct values of $P$.  

The dominant phase at point $P$ is determined by maximizing the entropy of the core black hole over the allowed region. 
We achieve this in two steps. First we recall the finding  of section \ref{sec:reg0}, that BPS black holes loose entropy by injecting additional charge into the gas. Therefore, the most advantageous configuration has no angular momentum in the gas and so the core black hole has $j=j_c$. 
In Figures \ref{fig:2comppossreg2} and \ref{fig:2comppossreg2b}) such core black holes lie on the part of the green curve that is on the boundary of the shaded patch. 

The next step is to maximize black hole entropy on the black hole sheet at constant value of $j$. In Section \ref{sec:bhsheet} we showed (below \eqref{bhentropy})
that the entropy decreases monotonically as the value of $q$ increases while keeping $j$ constant. Therefore, the
maximum entropy on the black hole sheet at constant value of $j$ is at the minimal $q$. This coincides with $\tilde q=0$ and is at the tip denoted $T$ in Figures \ref{fig:2comppossreg2} and \ref{fig:2comppossreg2b}. 

However, dependent on the total charge densities $(q,\tilde q, j)$ at the point $P$, the tip may or may not be consistent with the positivity constraints on the gas. In Figure \ref{fig:2comppossreg2} the tip is allowed, but in Figure \ref{fig:2comppossreg2b} it is not. This distinction divides region II into subregions that we call $II_a$ and $II_b$, respectively. In Figure \ref{fig:cohom} they are shaded orange and light blue. In both cases, candidate core black holes are on the black hole sheet and carry all the total angular momentum $j_c=j$. Positivity of the electric charge carried by the gas imposes $q_g \geq |\tilde q_g|$ and restricts the possible charges of the core black hole to the interior of the cone emanating from $P$ and drawn by dashed grey lines in Figures \ref{fig:2comppossreg2} and \ref{fig:2comppossreg2b}. The subsequent maximization of entropy along the black hole sheet favors the point nearest the tip $T$. 

The simplest is region IIa, where the tip is permitted. Then the core black hole has $\tilde q_c=0$, the total $\tilde q$ (if any) is carried by the gas. In contrast, for total charges $P$ 
in region $II_b$, the gas carries as much $\tilde q_g$ as possible, consistent with $q_g \geq |\tilde q_g|$. In this case the $(q_c, \tilde q_c)$ of the core black hole is obtained from the intersection between the black hole sheet and a $q_g = \tilde q_g$ (or $q_g=-\tilde q_g$) line from point $P$ (depending on whether $\tilde q>0$ or $\tilde q<0$). The dominant phase in region $II_b$ therefore contains the gas component which satisfies either $q_g = \tilde q_g$ or $q_g = -\tilde q_g$ depending on whether $\tilde q>0$ or $\tilde q <0$, respectively. That means gas component that carries either $q_{1g}$ only or $q_{2g}$ only.
 
\subsection{The Complete Supersymmetric Phase Diagram}
\label{sec:complete}
\begin{figure}[h]
    \centering
    \includegraphics[width=0.6\linewidth]{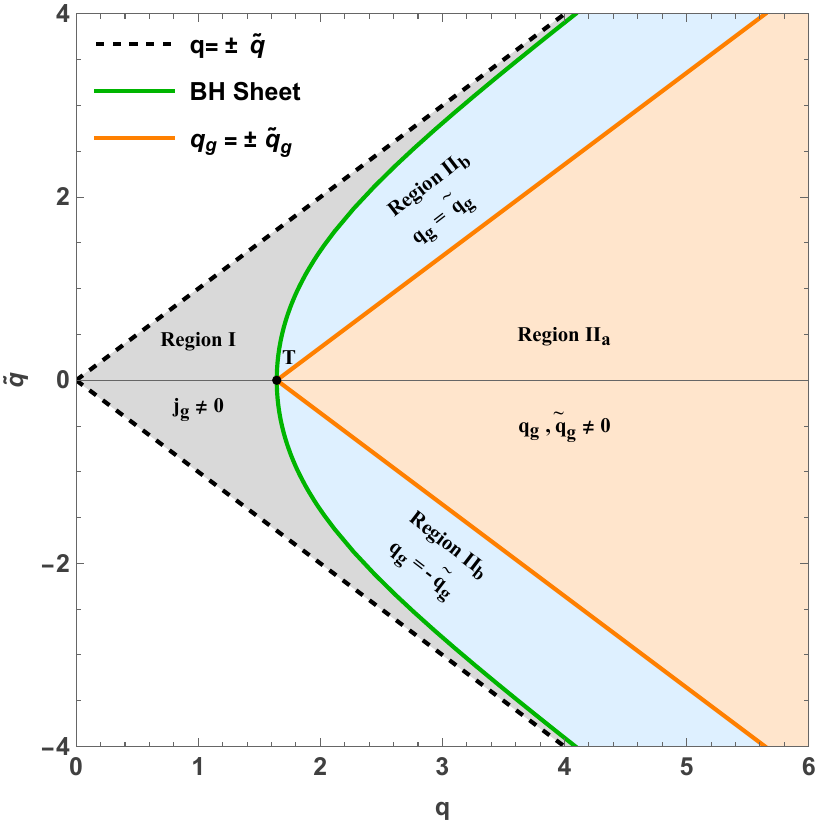}
    \caption{A plot of the constant $j$ cross-section of the microcanonical phase diagram of $AdS_4$ black holes with two charges. The black dashed lines form the boundaries of the allowed region in the charge space for the system as a whole. The black hole sheet that contains the pure black hole solutions is the green curve. Region I is the grey shaded area and represents configurations whose gas component carries only angular momentum. The core black holes in such configurations are the ones on the black hole sheet which are directly below the points at the same value of $(q,\tilde q)$. The orange lines form the boundaries of region $II_a$ in charge space where the dominant phase has a core black hole with $\tilde q=0$ at point $T$. Region $II_b$ is colored blue. Here the dominant phase has a gas component that carries either $q_{1g}$ only or $q_{2g}$ only.}
    \label{fig:cohom}
\end{figure}

The complete microcanonical phase diagram is shown in Figure \ref{fig:cohom}. The two component phases constructed in sections \ref{sec:reg1} and \ref{sec:reg2} fill the entire allowed region which is bounded by the three planes $q= |\tilde q|$ and $j=0$. It contains four distinct phases :
\begin{itemize}
    \item[(1)] {\it The black hole sheet}. A pure BPS black hole phase on the surface \eqref{sheetdef}. It is a green curve in Figure \ref{fig:cohom}.
    \item[(2)] {\it Region $I$} for total charges satisfying \eqref{reg1def}. A two component phase where the gas has angular momentum $j_g \neq 0$ and $q_g = \tilde q_g =0$. 
    \item[(3)] {\it Region $II_a$} for total charges that satisfy \eqref{reg2def} and inside the wedge $q_g = |\tilde q_g|$ from the point $T$. A two component phase with gas charges $j_g = 0$ and $q_g \neq  \tilde q_g $.
    \item[(4)] {\it Region $II_b$} for total charges that satisfy \eqref{reg2def} and outside the wedge $q_g = |\tilde q_g|$ from the point $T$. A two component phase with gas charges $j_g = 0$ and $q_g = \pm\tilde q_g $.
\end{itemize}

In Section \ref{sec:sci}, we utilize the supersymmetric phase diagram constructed here to predict that the superconformal index of the three-dimensional supersymmetric conformal field theory on the boundary exhibits a phase transition in the microcanonical ensemble. Specifically, we argue that, at leading order, the microcanonical superconformal index as a function of charges is well-approximated by the entropy of one of the two component configurations within certain regions of charge space. If this prediction can be independently verified through a direct computation of the index in the boundary theory, it would provide strong evidence for the existence of the novel two-component phases in $AdS_4 \times S^7$.

\subsection{Realization of the Gas Component}
\label{subsec:gas}

The phase diagram developed in Sections  
\ref{sec:reg0}--\ref{sec:complete}
was predicated on a simple hypothesis: there exists two component BPS configurations comprising a core black holes and a ``gas". The two component can coexist with their total energy and charges being a simple sum of contributions from the components, and the entropy entirely due to the core black hole. These assumptions were discussed in Section \ref{sec:princi}, without providing any specifics about the nature of the ``gas" component. In this subsection, we discuss arguments and supporting evidence for the principles. 

We will mostly rely on previous work in different, but closely related, settings. Two component solutions were proposed as endpoints of superradiant instabilities in non-supersymmetric black holes \cite{Kim:2023sig,Bajaj:2024utv,Choi:2024xnv}. Subsequently, it was conjectured that analogous solutions persist in the supersymmetric limit of Type IIB supergravity on $AdS_5 \times S^5$ \cite{Choi:2025lck}. Two component solutions in AdS$_3\times S^3$ were studied in \cite{Bena:2011zw,Larsen:2025jqo}. There is an important distinction between ``overrotation" and ``overcharging", even though the superradiant instability is formally similar in the two cases. We consider them in turn. 

Kerr black holes in AdS$_4$ are unstable to superradiance \cite{Cardoso:2004nk,Cardoso:2004hs}. A proposed endpoint of this instability is a two-component configuration consisting of a core black hole and a surrounding "gas" that orbits the black hole \cite{Kim:2023sig}. These solutions were termed ``Grey Galaxies". Their core black hole is non-supersymmetric and has angular velocity $(\Omega \ell = 1)$ determined by the onset of superradiance. The gas component saturates the BPS bound $E\ell = J$ by itself and revolves around the black hole in a disk within equatorial plane. The gas carries an $O(1)$ fraction of the angular momentum and energy of the full system but it is nonetheless dilute: it does not undergo gravitational collapse to form a second black hole. That is because large angular momentum pushes gas modes far away from the core black hole in the center. This large radial separation simplifies the construction of these solutions as the back reaction of the gas on the black hole and vice versa are suppressed by a factor of $G$. Analogous configurations were constructed in five dimensions, where, due to the presence of two independent angular momenta, the gas is spread over an $S^3\subset AdS_5$, rather than a single plane \cite{Bajaj:2024utv}. 

Reissner-Nordstr\"{o}m black holes in AdS are also unstable to superradiance. In this case, a proposed endpoint of the instability consists of a nonsupersymmetric core black hole at the superradiance bound $\Phi = 1$ and a dual giant graviton that carries electric charge, corresponding to angular momentum on an internal sphere \cite{Choi:2024xnv}. These solutions were termed ``Dual Dressed Black Holes''. 
In the context of black holes in $AdS_5 \times S^5$ the dual giant graviton is a $D3$-brane that wraps a $S^3\subset AdS_5$. On $S^5$ it is a point that caarries angular momentum. Our setting is $AdS_4 \times S^7$ where an analogous construction involves an $M2$-brane wrapping a $S^2 \subset AdS_4$ and is a point object in $S^7$ that rotates around a maximal circle. In empty AdS$_4$, the radial location where a dual giant can traverse a stable orbit is proportional to its electric charge $Q$ \cite{Grisaru:2000zn}. 
Therefore, when it carries macroscopic charges, it is far from the center of AdS$_4$, and so a very similar trajectory should be possible also in the presence of a core black hole. Moreover, the interaction between the two components is negligible at leading order. Black holes that are dressed by dual giant gravitons are more difficult to analyze than grey galaxies because they are localized, so backreaction can be substantial. 

In the supersymmetric limit, BPS black holes carry both angular momentum and electric charge. It is anticipated that the ``grey galaxy" and ``dual dressed black hole" constructions persist in this limit \cite{Choi:2025lck,deMelloKoch:2024pcs}. A core BPS black hole and a gas that are independently supersymmetric co-exist as a system at equilibrium. Interactions may not be literally negligible, there can be backreaction, but we expect that it is sufficiently benign that the energy of the total system is the sum of energies of the two components. 
Then the combined system will also be supersymmetric, because the BPS bound is a linear relation between energy and conserved charges.

\section{The Superconformal Index}\label{sec:sci}
\label{sec:supindex}
In this section we give a prediction for the microcanonical superconformal index in the dual 3D conformal field theory, by identifying the black hole configuration that has maximal entropy for given indicial charges. 

\subsection{Superconformal Index in 3D SCFT}\label{sec:3dsci}

In the grand canonical ensemble, the superconformal index is defined as the trace
\begin{equation}
    \mathcal I = {\rm Tr}\left[e^{-\beta \{\mathcal{Q},\mathcal{Q}^\dagger\}}e^{-\Phi_1^\prime (Q_1\ell) - \Phi_2^\prime (Q_2\ell) - \Phi_3^\prime (Q_3\ell) - \Phi_4^\prime (Q_4\ell)- \Omega^\prime J}\right]~, 
    \label{indprel}
\end{equation}
where the chemical potentials satisfy the constraint
\begin{equation}\label{indconst}
    \frac{1}{2}\left( \Phi_1^\prime + \Phi_2^\prime + \Phi_3^\prime + \Phi_4^\prime\right) - \Omega^\prime = 2\pi i ~.
\end{equation}
This constraint ensures that states which transform to each other by the action of the supercharge $\mathcal Q^{++++}_{-}$ with the quantum numbers $(Q_1\ell,Q_2\ell,Q_3\ell,Q_4\ell,J) = (\frac14,\frac14,\frac14,\frac14,-\frac12)$ do not contribute to the trace.\footnote{The electric charges are quarter-integers in units of the $AdS_4$ radius $\ell$. They are half-integers when written in terms of the $S^7$ radius because $\ell_{S^7} = 2 \ell$. } The trace \eqref{indprel} therefore only gets contributions from BPS states, ie. those that are annihilated by the supercharge $\mathcal{Q}$ (and the conjugate supercharge $\mathcal{Q}^\dagger$). The conserved charges of the BPS states satisfy the BPS relation
\begin{equation}
    \{\mathcal Q,\mathcal{Q}^\dagger\} = M\ell -( J + Q_1\ell + Q_2\ell + Q_3\ell + Q_4\ell) = 0 ~.
\end{equation}
We denote the degeneracy of the BPS states at given charges as $n(Q_1,Q_2,Q_3,Q_4,J)$ and its logarithm as the entropy $S(Q_1,Q_2,Q_3,Q_4,J)$. We expand the trace in the Index \eqref{indprel} as a sum over all BPS states in the theory:
\begin{equation}
\label{eqn:scindexdef}
    \mathcal I (\Phi_a^\prime) = \sum_{Q_a,J} n(Q_a,J) e^{2\pi i J }e^{-\Phi_1^\prime(Q_1\ell+\frac12J)-\Phi_2^\prime(Q_2\ell+\frac12J)-\Phi_3^\prime(Q_3\ell+\frac12J)-\Phi_4^\prime(Q_4\ell+\frac12J)}~,
\end{equation}
where $Q_a$ represents the four electric charges $(Q_1,Q_2,Q_3,Q_4)$. We substituted the constraint \eqref{indconst},
so the index is a function of only four chemical potentials $\Phi_a^\prime$. It follows that only four combinations of charges 
\begin{equation}
\label{eqn:Zadef}
    Z_a = Q_a\ell+\frac12J~,
\end{equation}
appear in the Boltzmann factors. We call them the \textit{indicial charges}. On the other hand, the summation \eqref{eqn:scindexdef} 
runs over all the five charges. To make it manifest that the index depends only on the indicial charges, 
we define quantities called the \textit{indicial entropy} and degeneracy which we denote by $S_{\mathcal I}$ and $n_{\mathcal I}$:
\begin{equation}\label{indent}
    n_{\mathcal I}(Z_a) = e^{S_{\mathcal I}(Z_a)} = \sum_{\underset{Q_a\ell+\frac12J=Z_a}{Q_a,J}} n(Q_a,J)e^{2\pi i J}~.
\end{equation}
The indicial degeneracy $n_{\mathcal I}$ depends only on the four indicial charges $Z_a$, because the summation in \eqref{indent}
is over all charges that lie on an \textit{indicial line} in the five dimensional charge space that is defined by the four linear equations \eqref{eqn:Zadef}.
Note that the indicial degeneracy can be a negative integer and hence the indicial entropy $S_{\mathcal I} $ can have an imaginary part. Using the definition of indicial entropy we can rewrite the superconformal index \eqref{indprel} as a sum over indicial charges $Z_a$:
\begin{equation}
    \mathcal{I}(\Phi_a^\prime) = e^{-\mathcal F(\Phi_a^\prime)} = \sum_{Z_a} e^{S_{\mathcal I}(Z_a) - \Phi_a^\prime Z_a}~,
\end{equation}
where $\mathcal F (\Phi_a^\prime)$ is the \textit{indicial free energy}.
The inverse of this formula can be obtained through an inverse Laplace transform 
\begin{equation}
\label{indext}
    e^{S_{\mathcal{I}}(Z_a)} \sim  \int d^4\Phi_a^\prime e^{-\mathcal{F}(\Phi_a^\prime) + \Phi_a^\prime Z_a}\approx \sum_{\Phi_a^{\prime\star}(Z_a)} e^{-\mathcal F(\Phi_a^{\prime\star}) + \Phi_a^{\prime\star}(Z_a)}~.
\end{equation}
The integral was approximated by a sum over saddle point values $\Phi_a^{\prime\star}$ of the chemical potentials. They are obtained by solving
\begin{align}
    \frac{\partial \mathcal F}{\partial \Phi_a^\prime} = Z_a   ~~~~~~~~~a=1,2,3,4~~~~~~, 
\end{align}
for $\Phi_a^\prime$ in terms of $Z_a$. The saddle point approximation is justified in the limit of large $N$ (or small $G$).

The summation in \eqref{indext} runs over all the solutions to the saddle point equations. In other words, the indicial entropy is a Legendre transform of the indicial free energy 
\begin{equation} \label{legtran}
    S_{\mathcal I}(Z_a) = {\rm ext}_{\Phi_a^\prime} \left\{-\mathcal F(\Phi_a^\prime)+\Phi_a^\prime Z_a  \right\}~.
\end{equation}
The extremization generally yields complex valued extrema. It was argued in \cite{Agarwal:2020zwm} that the unitarity of the CFT implies that the extrema appear in complex conjugate pairs, where the conjugate solution satisfies the constraint
\begin{equation}
    \frac{1}{2}\left( \Phi_1^\prime + \Phi_2^\prime + \Phi_3^\prime + \Phi_4^\prime \right)- \Omega^\prime = -2\pi i~.
\end{equation}
Let us denote the two solutions of the Legendre transform $S$ and $\bar S$. The saddle point value of the index can then be written as 
\begin{align}
    e^{S_{\mathcal I}(Z_a)} &= e^S+e^{\bar S}\\
    &= e^{{\rm Re}(S)} \cos ({\rm Im}(S))~.
\end{align}
The indicial entropy is therefore given by the formula
\begin{equation}
    S_{\mathcal I}(Z_a) = {\rm Re}(S) + \log\cos({\rm Im}(S))~.
\end{equation}
At leading order in $N$ ($\sim O(N^\frac32)$), we expect that the real part of $S$ (the Legendre transform of $\mathcal F$) equals the maximum entropy configuration on an indicial line defined by $Z_a$'s. The sign of the indicial degeneracy $n_{\mathcal I}(Z_a)$ also oscillates rapidly since the imaginary part of $S$ also scales like $N^\frac32$. The supergravity solutions are not sensitive to such rapid oscillations and therefore the bulk Lorentzian analysis of the index cannot be used to determine the imaginary part of the entropy ${\rm Im}(S)$.


Going forward, we restrict ourselves to the special case where two pairs of charges are set equal: $Q_1 = Q_3$ and $Q_2 = Q_4$. The chemical potentials satisfy analogous conditions: $\Phi_1^\prime = \Phi_3^\prime$ and $\Phi_2^\prime = \Phi_4^\prime$. In this sector the superconformal index \eqref{eqn:scindexdef} reduces to 
\begin{equation}
    \mathcal I (\Phi_1^\prime,\Phi_2^\prime) = e^{-\mathcal F(\Phi_1^\prime,\Phi_2^\prime)} = \sum_{Z_1,Z_2}e^{S_{\mathcal I}(Z_1,Z_2) - 2\Phi_1^\prime Z_1 - 2\Phi_2^\prime Z_2}~.
\end{equation}
We further change variables
\begin{align}
     \Phi^\prime &= \Phi_1^\prime+\Phi_2^\prime~,\\
     \tilde \Phi^\prime &= \Phi_1^\prime-\Phi_2^\prime~,
\end{align}
in the definition of the superconformal index and obtain 
\begin{align}
    \mathcal I (\Phi^\prime, \tilde \Phi^\prime) = e^{-\mathcal F(\Phi^\prime,\tilde \Phi^\prime)} = \sum_{\zeta,\tilde\zeta} e^{k\left(s_{\mathcal I}(\zeta,\tilde\zeta) - \Phi^\prime \zeta - \tilde\Phi^\prime \tilde\zeta\right)} ~. 
\end{align}
The constant $k\sim N^{\frac3 2}$ was defined in \eqref{eqn:kdef} to normalize the densities introduced in (\ref{q1q1def}-\ref{linmin}).
The two indicial charge densities $\zeta$ and $\tilde\zeta$ are related to $Z_1$ and $Z_2$ as
\begin{align}
\label{eqn:zeta1def}
    \zeta &= k^{-1} (Z_1+Z_2) = q + j ~,\\
    \tilde\zeta &= k^{-1} (Z_1 - Z_2) = \tilde q ~.
    \label{eqn:zeta2def}
\end{align}

In the rest of this section, we compute the indicial entropy density $s_{\mathcal{I}}(\zeta,\tilde\zeta)$ using the phase diagram constructed in section \ref{sec:2comp}. We perform the computation by making the following assumption: the indicial entropy defined through the rapidly oscillating sum \eqref{indent} over configurations on an indicial line in charge space is dominated by the entropy of the configuration with the largest entropy.\footnote{This statement was named the `Unobstructed Saddle Conjecture' in \cite{Choi:2025lck}.}

\subsection{The Indicial Lines}\label{sec:indline}

\begin{figure}[t]
    \centering
    \includegraphics[width=0.5\linewidth]{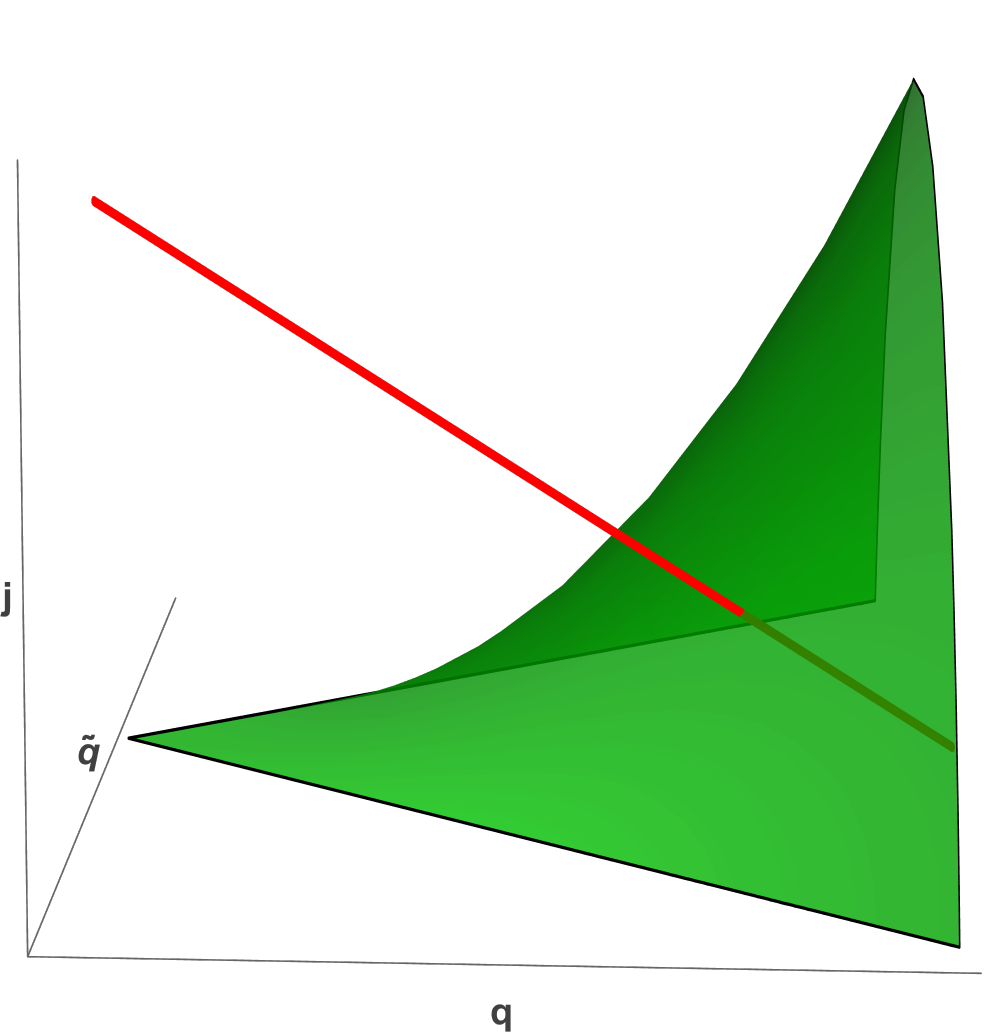}
    \caption{The 3 dimensional depiction of the black hole sheet (green surface) from Figure \ref{fig:BHsheet}, now augmented with an indicial line (red) that intersects the black hole sheet at one point. As we move on the line in the direction of increasing $q$, the value of $j$ decreases so that $\zeta = q+j$ remains constant.}
    \label{fig:indeline}
\end{figure}

An indicial line in the three dimensional charge space is specified by the parameters $(\zeta,\tilde\zeta)$. 
Their definitions (\ref{eqn:zeta1def}-\ref{eqn:zeta2def}) and positivity of charges $q\geq |\tilde q|$, $J\geq 0$ impose
\begin{equation}\label{indlimit}
    \zeta \geq |\tilde\zeta|~,
\end{equation}
for indicial lines that cross the physical region. Saturation of \eqref{indlimit} is a degenerate special case where the index line intersects the physical region only at a single point where $q=|\tilde q|=|\tilde\zeta|$ and $j=0$. In other cases the index line $(\zeta,\tilde\zeta)$ is physical for a finite range of $q$ that parametrizes the position along the indicial line. Generic index lines intersect the black hole sheet exactly once, as illustrated in Figure \ref{fig:indeline}. In the rest of this section, we will restrict ourselves to the case when $\tilde\zeta>0$.   

\begin{figure}[t]
    \centering
    \includegraphics[width=0.5\linewidth]{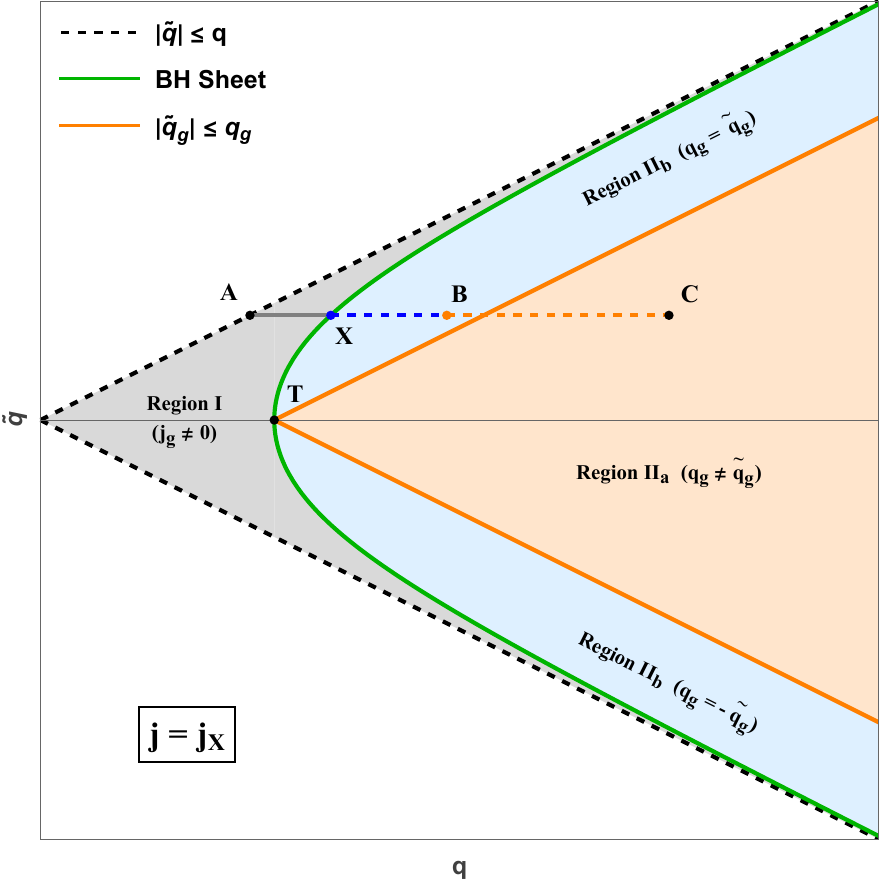}
    \caption{The constant $j$ phase diagram at $j=j_X$. A projection of the indicial line is horizontal on the plot, because it has constant $\tilde q$. We keep only the line segment between $A$ and $C$ that intersects the physical region. The grey (solid line) part of the line segment ($AX$) lies in Region $I$ and has $j>j_X$ (above the plane), the blue (dashed) part of the line segment ($XB$) lies in Region $II_b$ and has $j<j_X$ (below the plane), and the orange (dashed) part of the line segment ($BC$) lies in Region $II_a$ and also has $j<j_X$. The point $B$ where the indicial line transitions between Region $II_b$ and Region $II_a$ is at $j=j_B<j_X$ where the orange region has shifted to the left, so the projection of $B$ to $j=j_X$ is not on the boundary of Region $II_a$.}
    \label{fig:constjxind}
\end{figure}

\begin{figure}[t]
    \centering
    \includegraphics[width=0.5\linewidth]{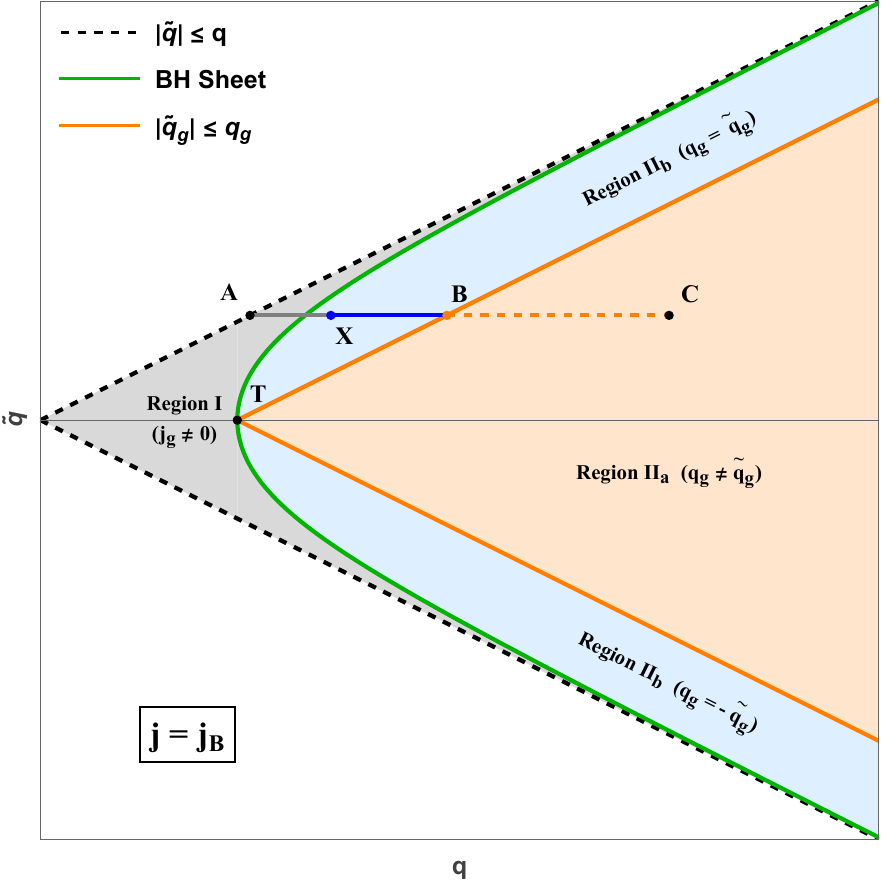}
    \caption{The constant $j$ phase diagram at $j=j_B$ with a projection of the indicial line. We focus on the line segment between points $A$ and $C$ that intersects with the physical region. The portion of the line segment to the left of point $B$ lies at $j>j_B$ which above the plane and drawn as a thick line, whereas the portion to the right of $B$ lies at $j<j_B$ which is below the plane and drawn as a dashed line. The point $B$ lies at the transition from Region $II_b$ and $II_a$ at $j=j_B$. The point $X$ is at $j=j_X<j_B$ where the green black hole sheet has moved right from its position at $j=j_B$. }
    \label{fig:constjBind}
\end{figure}

The projection of indicial lines onto $q-\tilde q$ plots like those drawn in Figures \ref{fig:constj}--\ref{fig:cohom} always gives a horizontal line, because $\tilde q=\tilde\zeta$ is kept constant. Restricting the indicial line to the physical region where charges are positive, we can take it to begin at the $q=|\tilde q|$ plane. This is point $A$ in Figures \ref{fig:constjxind} and \ref{fig:constjBind}. As $q$ increases, the index line continues through the grey shaded region where the gas component carries only $j$. At some point $X$, it intersects the black hole sheet indicated the green curve. For larger $q$, 
the index line enters the blue shaded region where the gas component carries either $q_1$ or $q_2$ charge, but not both. As $q$ increases even more, at point $B$ the indicial line crosses into the orange region, where the two component configurations consist of a core black hole with $\tilde q=0$ (and so $q_1=q_2$) and a gas with both charges ($q_{1,2} \neq 0$). In the special case $\tilde\zeta=0$, the index line avoids the blue region and goes directly to the region shaded orange.

All of our plots of the $q-\tilde q$ plane have fixed value of $j$. However, as $q$ increases along an index line the value of $j$ decreases, and so the line moves to $q-\tilde q$ planes with smaller $j$. 
Figure \ref{fig:constjxind} depicts the $j=j_X$ slice which contains the intersection point $X$. The index line is dashed to the right of $X$, where $j$ has a smaller value. At these lower $j$, region IIb (shaded with orange) is larger, and so the point $B$ where it is reached by the index line is at smaller value of $q$. Therefore, in Figure \ref{fig:constjxind}, $B$ is strictly to the left of the orange line that indicates the boundary of region IIb when $j=j_X$. Conversely, Figure \ref{fig:constjBind} visualizes the phase diagram at constant $j=j_B$. The index line is solid to the left of $B$, because it is at $j>j_B$ there. At such values of $j$ the black hole sheet has moved further to the right. In the projection to $j=j_B$ is appears that the point $X$ is in region IIb but in fact it is on the green interface between regions I and IIb. 
In both of Figures \ref{fig:constjxind} and \ref{fig:constjBind} the index line continues towards smaller $j$ as it moves deeper into region IIb. The line segment ends at the point $C$ because it exits the physical region on the $j=0$ plane.

A brief summary of the discussion above: the points $A,X,B$ and $C$ split a generic index line into three segments -- $i_I$, $i_{II_a}$, and  $i_{II_b}$ -- which lie in regions $I,II_a$ and $II_b$ respectively. 
The line segment $i_I$ lies between points $A$ and $X$, $i_{II_b}$ lies between points $X$ and $B$, and $i_{II_a}$ lies between points $B$ and $C$. In the following we identify the coordinates of the joining points $A, X, B$, and $C$ as functions of the indicial charges $(\zeta,\tilde\zeta)$.

The leftmost point $A$ on segment $i_I$ is vertically (in $j$ direction) above a point on the $q=|\tilde q|$ line. The coordinates of $A$ 
follow from the definition of the indicial line (\ref{eqn:zeta1def}-\ref{eqn:zeta2def})
and the condition that $A$ also lies on the $q=|\tilde q|$ plane: 
\begin{align}
    \zeta &= q + j~,\\
    \tilde\zeta &= \tilde q~,\\
    q&=\tilde q~. 
\end{align}
The solutions for the coordinates of $A$ is $(q,\tilde q,j)=(\tilde\zeta,\tilde\zeta,\zeta-\tilde\zeta)$. 

The point $X$ is on the black hole sheet
and the common endpoint of segments $i_I$ and $i_{II_b}$. The definition of the indicial line (\ref{eqn:zeta1def}-\ref{eqn:zeta2def})
and the condition \eqref{sheetdef} that the point is on the black hole sheet give:
\begin{align}
    \zeta &= q_X + j_X~,\\
    \tilde\zeta &= \tilde q_X~,\\
    j_X &= q_X \left(-1+\sqrt{1+4\left(q_X^2-\tilde q_X^{2}\right)}\right)~,
\end{align}  
for $(q_X,\tilde q_X,j_X)$ as functions of $(\zeta,\tilde\zeta)$ which parameterize the indicial line. Eliminating $j_X$ and $\tilde q_X$, we get a quadratic for $q^2_X$
\begin{equation}
    4 q_X^4 +(1-4\tilde\zeta^2)q_X^2 - \zeta^2 =0~,
\end{equation}
with the solutions 
\begin{equation}
    q_X^2 = \frac18 \left(4\tilde\zeta^2-1 + \sqrt{(1-4\tilde\zeta^2)^2+16\zeta^2}\right)~. \label{q2sol}
\end{equation}
Positivity $q_X^2>0$ determined the branch. The coordinates $(q_X,\tilde q_X,j_X)$ of the point $X$ become
\begin{align}
    q_X &= \frac{1}{2\sqrt 2} \sqrt{4\tilde\zeta^2-1 + \sqrt{(1-4\tilde\zeta^2)^2+16\zeta^2}}~,\label{qxcoord} \\
    \tilde q_X &= \tilde\zeta~, \label{qtildexcoord}\\
    j_X &= \zeta - \frac{1}{2\sqrt 2} \sqrt{4\tilde\zeta^2-1 + \sqrt{(1-4\tilde\zeta^2)^2+16\zeta^2}}~.
\end{align}
At the point $X$ we are simply considering a pure BPS black hole, albeit as function of indicial charges. 
After inserting \eqref{q2sol} and taking $\tilde q = \tilde\zeta$, the formula for entropy \eqref{bhentropy} as a function of $(q,\tilde q)$:
\begin{equation}
    s(q,\tilde q) = \pi \left(-1+\sqrt{1+4(q^2-\tilde q^2)}\right)~,
\end{equation}
gives,
\begin{equation}\label{indxent}
    s_X(\zeta,\tilde\zeta) = \pi \left(-1+ \sqrt{\left(\frac12 - 2 \tilde\zeta^2\right)+ \sqrt{\left(\frac12 - 2 \tilde\zeta^2\right)^2+4 \zeta^2}}\right)~.
\end{equation}

Point B is the common endpoint of the segments $i_{II_a}$ and $i_{II_b}$. At B the gas carries all $\tilde q_g=\tilde q_B$ charge and it saturates $q_g\geq |\tilde q_g|$, so the core black hole has charge assignments 
$(q_c,\tilde q_c,j_c)=(q_B-\tilde q_B,0,j_B)$.
The definition of the indicial line (\ref{eqn:zeta1def}-\ref{eqn:zeta2def}) and the condition that the core black hole satisfy the constraint \eqref{sheetdef} then give
\begin{align}
    \zeta &= q_B + j_B~,\\
    \tilde\zeta &= \tilde q_B~,\\
    j_B &= (q_B-\tilde q_B)\left(-1+\sqrt{1+4 (q_B-\tilde q_B)^2}\right)~.\label{2ndpoint}
\end{align}
The charges at the point $B$ become:
\begin{align}
    q_B &= \tilde\zeta + \frac{1}{2\sqrt2}\sqrt{-1+\sqrt{1+16(\zeta-\tilde\zeta)^2}}~, \label{qbcoord}\\
    \tilde q_B &= \tilde\zeta~,\\
    j_B &= (\zeta-\tilde\zeta)-\frac{1}{2\sqrt2}\sqrt{-1+\sqrt{1+16(\zeta-\tilde\zeta)^2}}~.
\end{align}

Finally, the other endpoint $C$ of line segment $i_{II_a}$ is located at $j=0$, so the coordinates of this point are given by $(q,\tilde q, j)=(\zeta,\tilde\zeta,0)$. 

\subsection{Entropy Along an Indicial Line}\label{sec:entind}

In the previous subsection we divided the indicial line with parameters $(\zeta,\tilde\zeta)$ into three segments $i_I$, $i_{II_b}$ and $i_{II_a}$. We now proceed to compute the entropy on each line segment, as function of the parameter $q$ that gives the position along the line.

The line segment $i_I$ lies entirely in Region I, the grey region of Figures \ref{fig:constjxind} and \ref{fig:constjBind}. Here the gas carries only angular momentum $j_g$, so the charges $(q_c,\tilde q_c)$ of the core black hole are the same as the total charges $(q,\tilde q)$ of the two component system.
Therefore, the entropy on line segment $i_I$ is simply the entropy \eqref{bhentropy} of a BPS black hole with fixed $\tilde q=\tilde\zeta$:
\begin{equation}\label{iIent}
    s(q,\zeta,\tilde\zeta) =\pi \left(-1+\sqrt{1+4(q^2-\tilde \zeta^2)}\right) ~.
\end{equation}
The leftmost point on the line segment $i_I$ is $A$. It has the smallest value of $q$ and lies on the surface $q=|\tilde q|=\tilde\zeta$ so the entropy \eqref{iIent} vanishes at $A$. As we move to the right on the line by increasing the value of $q$, the entropy increases monotonically until we reach the intersection point $X$, where the entropy takes the value $s_X$ given in \eqref{indxent}. 

The line segment $i_{II_b}$ lies entirely in the blue region of Figures \ref{fig:constjxind} and \ref{fig:constjBind}. In this region, the gas component carries either $q_{1g}$ only (when $\tilde q>0$) or $q_{2g}$ only (when $\tilde q<0$) and $j_g=0$. As we move from one end of the line segment (point $X$) to the other (point $B$), the corresponding core black holes trace a path on the black hole sheet. We call this path the \textit{shadow} of the index line. 
We seek the entropy along the indicial line as a function of a parameter  \textit{on} the indicial line. To obtain this, we first solve the equations
\begin{gather}
    q+j=\zeta~, \label{reg2beq1}\\
    \tilde q = \tilde\zeta~, \label{reg2beq2}\\
    q-q_c = \tilde q - \tilde q_c~, \label{reg2beq3}\\
    j = q_c\left(-1+\sqrt{1+4\left(q_c^2-\tilde q_c^{2}\right)}\right)~,\label{reg2beq4}
\end{gather}
for the four variables $(j,\tilde q,q_c,\tilde q_c)$ as a function of $(q,\zeta,\tilde\zeta)$. The first two equations define the indicial line. The third equation implies that the charges of the gas satisfy $q_g = \tilde q_g$. The last equation ensures that the core black hole is on the black hole sheet. Note that we have used the fact that the angular momentum of the core black hole $j_c$ is equal to the total angular momentum $j$ while writing the last equation. The entropy on line segment $i_{II_b}$ is then obtained by substituting the charges of the core black holes $j(q,\zeta,\tilde\zeta)$ and $q_c(q,\zeta,\tilde\zeta)$ in the BPS entropy formula: 
\begin{equation}\label{reg2bent}
    s(q,\zeta,\tilde\zeta) = \pi \frac{j(q,\zeta,\tilde\zeta)}{q_c(q,\zeta,\tilde\zeta)} ~.
\end{equation}

The final leg of the indicial line is the segment $i_{II_a}$ which lies entirely in region $II_a$. The two component configurations in this region of the charge space have core black holes with $\tilde q_c=0$ and $j_c=j$. Therefore, the shadow of the line segment $i_{II_a}$ is at the intersection between the $\tilde q=0$ plane and the black hole sheet. 
The parameters of the shadow $(j,q_c)$ are given as function of $(q,\zeta,\tilde\zeta)$ through
\begin{gather}
    j=\zeta-q~, \label{reg2ajj}\\
    j=q_c\left(\sqrt{1+4q_c^2}-1\right)~.\label{reg2aj}
\end{gather}
The entropy of the configurations on the line segment $i_{II_a}$ then follows by substituting the function $q_c(q,\zeta,\tilde\zeta)$ and $\tilde q_c = 0$ in the entropy formula \eqref{bhentropyqqt} for the core black holes
\begin{equation}\label{bhent2a}
    s(q_c) = \pi \left(\sqrt{1+4q_c^2}-1\right) ~.
\end{equation}
As we increase $q$ along the line segment $i_{II_a}$, 
the value of $j$ decreases, according to \eqref{reg2ajj}, and then \eqref{reg2aj} shows that $q_c$ decreases as well. We conclude that the entropy \eqref{bhent2a} decreases monotonically as we move along $i_{II_a}$ between points $B$ and $C$. The maximum entropy configuration on line segment $i_{II_a}$ is at point $B$ and the minimum $s=0$ is at $C$.

\begin{figure}
    \centering
    \includegraphics[width=0.7\linewidth]{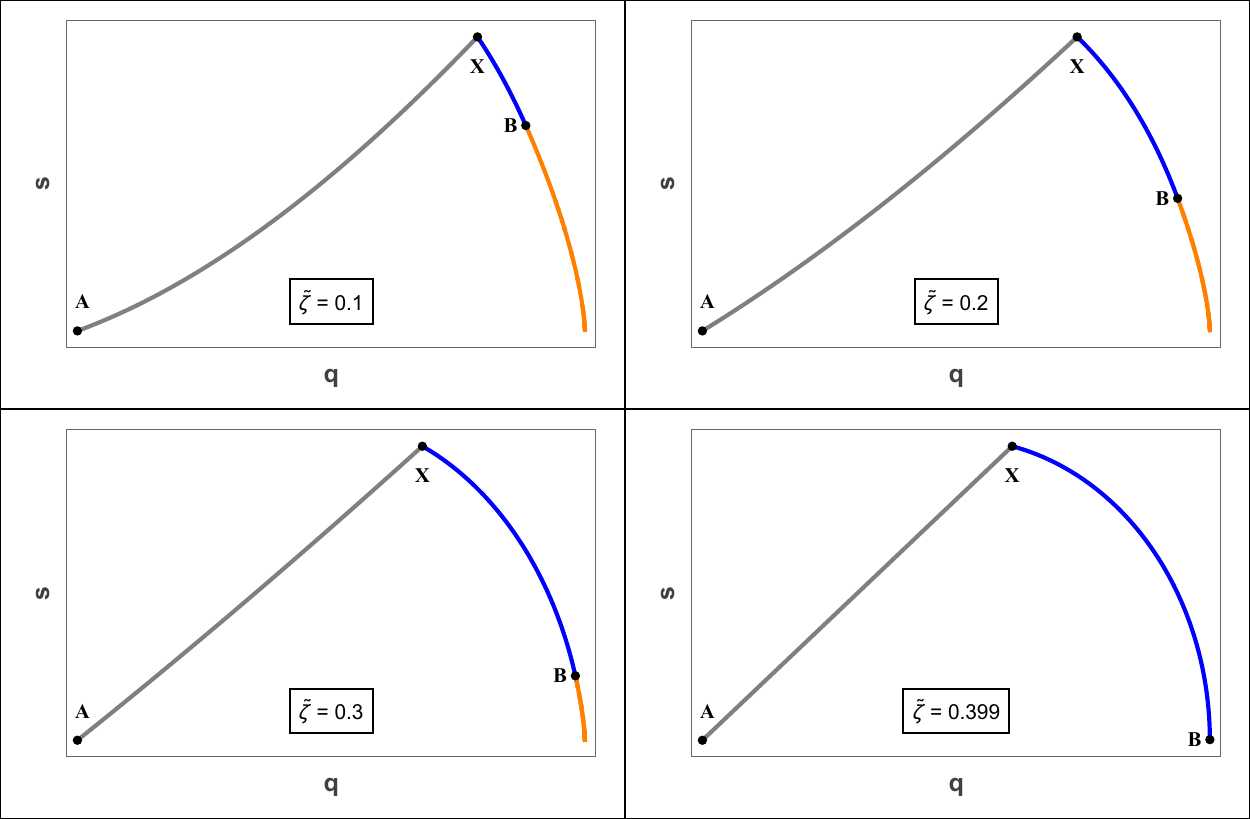}
    \caption{The entropy of two component configurations along index lines with fixed $\zeta = 0.4 <\frac{1}{2}$ and various values of $\tilde\zeta$ satisfying $0<\tilde\zeta<\zeta$. The configuration with the maximum entropy is always the pure BPS black hole at point $X$ where the indicial line transitions from Region $I$ to Region $II$. This is where  grey and blue curves meet in the plots.}
    \label{fig:indexlinebelow}
\end{figure}

\begin{figure}
    \centering
    \includegraphics[width=0.7\linewidth]{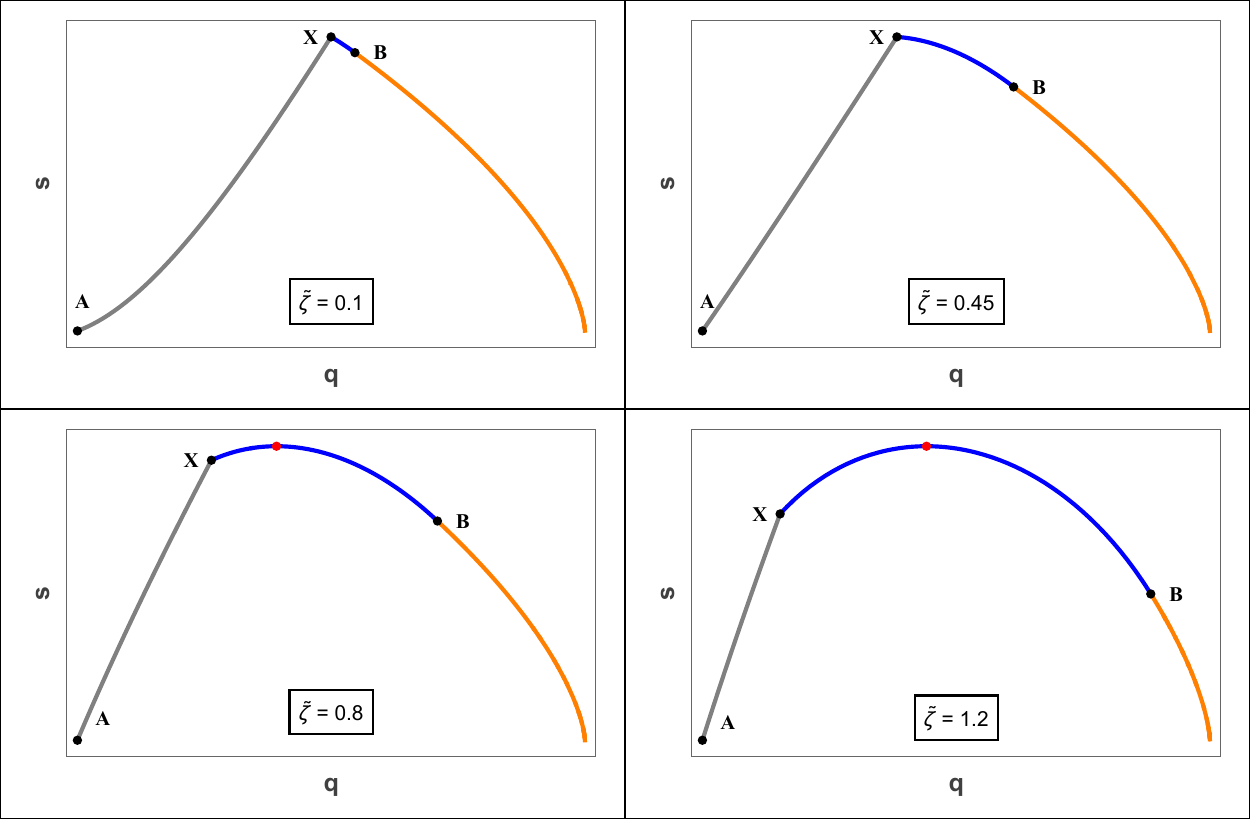}
    \caption{The entropy of two component configurations along index lines with fixed $\zeta = \frac32 >\frac{1}{2}$ and various values of $\tilde\zeta$ satisfying $0<\tilde\zeta<\zeta$. For $\tilde\zeta<\frac12$ the maximum is at $X$, as in Figure \ref{fig:indexlinebelow}, but, when $\tilde\zeta>\frac12$, there is a novel maximum marked by a red dot. In the text we show that the dominant configuration is in a two component phase for any $\tilde\zeta>\frac12$.}
    \label{fig:indexlineabove}
\end{figure}

The entropy of two component configurations on a sample of indicial lines are plotted in Figures \ref{fig:indexlinebelow} and \ref{fig:indexlineabove}. We observe the following qualitative features:
\begin{itemize}
    \item The entropy vanishes at both end points of all indicial lines. The core black holes have zero entropy at the end points because they have $j=0$ and lie on the $q=\pm \tilde q$ lines.  
    \item For indicial lines with $\zeta<\frac{1}{2}$ (Figure \ref{fig:indexlinebelow}), the entropy is always maximized at point $X$, where the indicial line intersects the black hole sheet. 
    \item For indicial lines with $\zeta>\frac{1}{2}$ (Figure \ref{fig:indexlineabove}) and $|\tilde\zeta|<\frac{1}{2}$, the entropy is still maximized at point $X$, and so the index is again dominated by a pure BPS black hole.
    \item For $\zeta>\frac{1}{2}$ (Figure \ref{fig:indexlineabove}) and $\tilde\zeta>\frac{1}{2}$, the entropy reaches its maximum at some point on the line segment $i_{II_b}$. This corresponds to a two component configuration. 
\end{itemize}
In the next subsection we show that all index lines with
$\tilde\zeta\leq \frac{1}{2}$ have maximal entropy on the black hole sheet, but those with $\tilde\zeta>\frac{1}{2}$ have maximum in the two component phase. The dependence on $\zeta$ enters through the requirement \eqref{indlimit} that all indicial lines satisfy $\zeta\geq |\tilde\zeta|$. Therefore, the new phases with $\tilde\zeta>\frac{1}{2}$ are only possible if
we also have $\zeta>\frac{1}{2}$.

\subsection{Maximizing Entropy Along the Indicial Line} \label{sec:maxsegb}

Figures \ref{fig:indexlinebelow} and \ref{fig:indexlineabove} suggest that
the microcanonical index undergoes a phase transition. The pure BPS black hole dominates the index in most of the representative plots but, when $\tilde\zeta$ exceeds a critical value, a two component configuration gives the largest contribution. In this subsection we derive the locations of the maxima from black hole thermodynamics.

In Section \ref{sec:entind} we already showed that, as we move along any indicial line $(\zeta,\tilde\zeta)$, the entropy increases monotonically on the line segment $i_{I}$ $(AX)$ and decrease monotonically on $i_{II_a}$ ($BC$). Therefore, the maximum entropy is always on the middle line segment $i_{II_b}$ $(XB)$. We will see below that it is at $X$, corresponding to a pure BPS black hole, 
when $\tilde\zeta$ is below the critical value $\tilde\zeta=\frac{1}{2}$. When $\tilde\zeta>\frac{1}{2}$, it is somewhere in the bulk of the line segment $i_{II_b}$. The maximum approaches $B$ as $\tilde\zeta\to\infty$, but it never reaches it. 

The motion along an index line corresponds to increasing value of $q$. We first identify the point $q$ where the entropy is extremized and then compute the extremum value of the entropy. On the black hole sheet, the black hole entropy is:
\begin{equation}\label{reg2bentf}
    s(q_{1c},q_{2c}) = \pi\left(-1+\sqrt{1+16 q_{1c}q_{2c}}\right)~.
\end{equation}
Here we have reverted back to the variables $q_{1,2}$ which are related to $q,\tilde q$ through \eqref{linpl} and \eqref{linmin}. The charges of the core black holes in a two component solution are marked with a subscript ${}_c$. They satisfy the black hole sheet constraint \eqref{sheetdef} in the form
\begin{equation}
    j_c = (q_{1c}+q_{2c}) \left(-1+\sqrt{1+16 q_{1c} q_{2c}}\right)~. \label{sheetconst}
\end{equation}
The total charges lie on the indicial line and do not carry the ${}_c$ subscript.

\begin{figure}[H]
    \centering
    \includegraphics[width=0.5\linewidth]{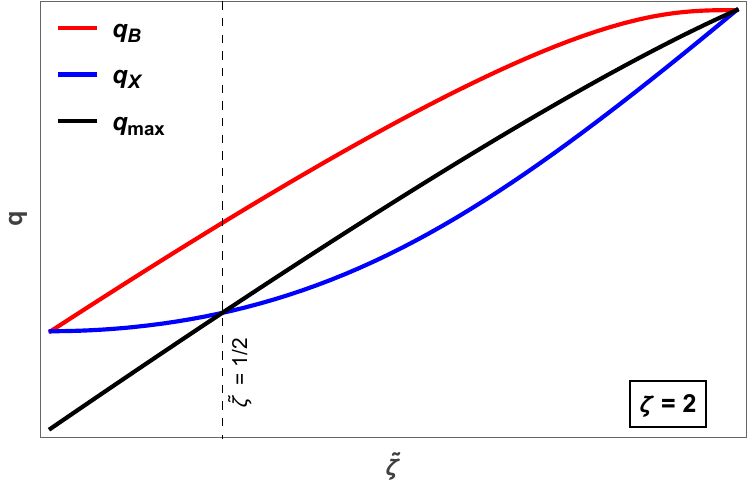}
    \caption{The $q$ coordinates of the points $X$, $B$, and the candidate maximum \eqref{indmaxq}, for the range $0\leq \tilde\zeta\leq\zeta$. The line segment $i_{II_b}$ lies between the points $X$ and $B$. The proposed maximum lies within this range exactly when $\tilde\zeta>\frac12$. The plot is for the line with $\zeta = 2$ but representative of all $\zeta > \frac{1}{2}$. 
    }
    \label{fig:qbqmaxqx}
\end{figure}

As we move along the index line, the core black holes move along the shadow on the black hole sheet. The motion along the $i_{II_b}$ segment of the indicial line, the blue line segment between points $X$ and $B$ in Figures \ref{fig:constjxind} and \ref{fig:constjBind},
is described by (\ref{reg2beq1}--\ref{reg2beq3}). Eliminating the parameter $q$ that gives the position along the index line, we find
\begin{equation}\label{reg2bconst}
    q_c - \tilde q_c + j= \zeta - \tilde\zeta~. 
\end{equation}
The gas carries $j_g=0$, so the total $j$ is carried by the core black hole $j=j_c$. The direction of the shadow on the black hole sheet then becomes\footnote{This condition is valid in the region $II_b$ above the $\tilde q =0$ axis. The analysis for the region $II_b$ below the $\tilde q=0$ axis is analogous, but the additional constraint is $$2dq_{1c}+dj_c=0~.$$}
\begin{align}
    2dq_{2c}+dj_c=0 ~\implies ~ dq_{2c} \left(2+\frac{\partial j_c}{\partial q_{2c}}\right)+ \frac{\partial j_c}{\partial q_{1c}} dq_{1c} = 0~, \label{reg2bcondn}
\end{align}
with derivatives of $j_c$ computed from \eqref{sheetconst}.
The change in the entropy due to charges shifting in the direction of the shadow along the sheet is 
\begin{equation}
\label{eqn:dsIIb}
ds = \frac{\partial s}{\partial q_{1c}} dq_{1c} + \frac{\partial s}{\partial q_{2c}} dq_{2c}= \left(-\frac{\partial s}{\partial q_{1c}}\left(\frac{2+\frac{\partial j_c}{\partial q_{2c}}}{\frac{\partial j_c}{\partial q_{1c}}}\right)+\frac{\partial s}{\partial q_{2c}}\right) dq_{2c}~, 
\end{equation}
where we inserted the condition \eqref{reg2bcondn}. This gives the extremization condition on the entropy:
\begin{equation}
    \frac{\partial s}{\partial q_{2c}} \frac{\partial j_c}{\partial q_{1c}} = \frac{\partial s}{\partial q_{1c}}\left(2+ \frac{\partial j_c}{\partial q_{2c}}\right)~,
\end{equation}
which, for the entropy formula \eqref{reg2bentf}, reduces to
\begin{equation}
\label{eqn:q1coverq2c}
    \frac{q_{1c}}{q_{2c}} = \frac{2+\frac{\partial j_c}{\partial q_{2c}}}{\frac{\partial j_c}{\partial q_{1c}}}~.
\end{equation}
Inserting the black hole charge constraint \eqref{sheetconst}, we arrive at the simple solution
\begin{equation}
    q_{1c} - q_{2c} = \frac12 = \tilde q_c~, \label{indmaxcond}
\end{equation}
after a simple computation.
We identify the location $q_{max}$ of the extremum on the indicial line $(\zeta,\tilde\zeta)$ by substituting the result \eqref{indmaxcond} in the equations \eqref{reg2beq1} - \eqref{reg2beq4}. 
This gives
\begin{gather}
    q_{max}  = q_c + \tilde\zeta - \frac{1}{2}~,\\
    j_c  =  \zeta-q = q_c(2q_c-1)~, \label{indmaxjc}
\end{gather}
and so
\begin{align}
q_c &= \frac{1}{2}\sqrt{1 + 2 (\zeta - \tilde\zeta)}~, \label{indmaxqc}\\
q_{max} &= \tilde\zeta + \frac12\left(\sqrt{1+2(\zeta-\tilde\zeta)}-1\right) ~. \label{indmaxq}
\end{align}
The point $q_{\rm max}$ \eqref{indmaxq} specifies the position of the extremum along the indicial line $(\zeta,\tilde\zeta)$. The parameters $(q_c,\tilde{q}_c,j_c)$ of the corresponding core black hole are given in \eqref{indmaxqc}, \eqref{indmaxcond} and \eqref{indmaxjc}  respectively. We have marked the $q=q_{max}$ points on the indicial lines in Figure \ref{fig:indexlineabove} (red points). Substituting the expression \eqref{indmaxqc} for $q_c$ and $\tilde q_c = 1/2$ in 
\begin{equation}
    s(q_c,\tilde q_c) = \pi \left(-1+\sqrt{1+4(q_c^2-\tilde q_c^2)}\right)~. 
\end{equation}
we obtain the entropy at the maximum: 
\begin{equation}\label{segbmax}
    s_{max}(\zeta,\tilde\zeta) = \pi \left(-1+\sqrt{1+2(\zeta-\tilde\zeta)}\right)~.
\end{equation}
For consistency, the point $q=q_{max}$ \eqref{indmaxq} must lie between the points $X$ and $B$, because it was identified by using formulae that apply only there. 
The expressions for the value of $q$ at points $X$ and $B$ are given in \eqref{qxcoord} and \eqref{qbcoord}, respectively. In Figure \ref{fig:qbqmaxqx}, we have plotted these expressions as function of $\tilde\zeta$ in the range $0<\tilde\zeta<\zeta$ with $\zeta=2$. We see that 
$q_{max} < q_B$ for all $\tilde\zeta$, so $q_{\rm max}$ never moves beyond $B$. In contrast, $q_X<q_{max}$ only when $\tilde\zeta > \frac12$, so only then is the candidate $q_{max}$ \eqref{indmaxq} legitimate. For $\tilde\zeta < \frac12$, the entropy is maximized at $X$. Although the plot considers the specific value $\zeta = 2$, we have verified using \texttt{Mathematica} that this value is representative of all $\zeta>\frac{1}{2}$.

\begin{figure}
    \centering
    \includegraphics[width=\linewidth]{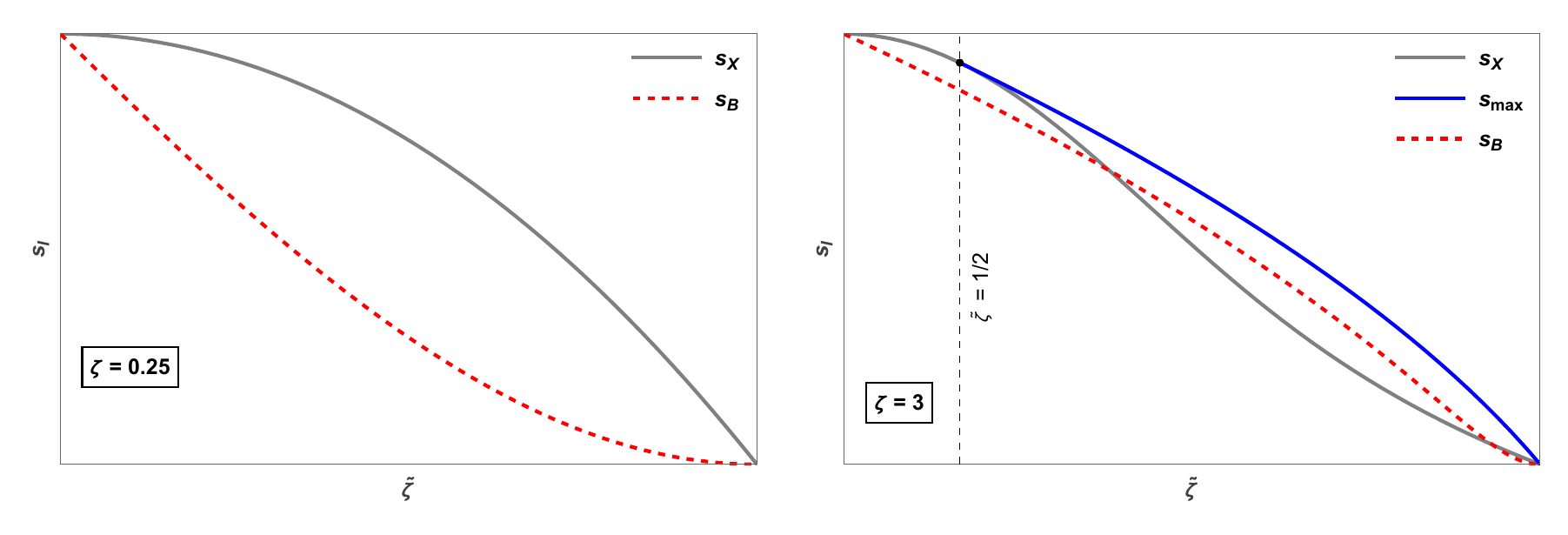}
    \caption{The plots show the entropies of the three supersymmetric configurations at the points $q=q_X$, $q=q_{max}$, and $q=q_B$ on the indicial line $(\zeta,\tilde\zeta)$ for the range $\tilde\zeta\in(0,\zeta)$ with fixed $\zeta$. The left side plot has $\zeta= 0.25<\frac12$ and the right side plot has $\zeta = 3> \frac12$. The two component phase that dominates the microcanonical ensemble when $\tilde\zeta>\frac12$ is the blue curve on the right side where $\zeta>\frac{1}{2}$.
    }
     \label{fig:indphase}
\end{figure}

The microcanonical index $s_{\mathcal I}(\zeta,\tilde\zeta)$ is approximated by the maximum entropy configuration along a given indicial line $(\zeta,\tilde\zeta)$. Three points along the line should be considered: 
the pure BPS black hole at point $X$ with entropy given by \eqref{indxent}, the two component configuration at point $B$ with entropy given by \eqref{bhent2a} (where the function $q_c(q,\zeta,\tilde\zeta)$ is evaluated at $q=q_B$), and the extremum \eqref{indmaxq} in the bulk of tthe line segment $i_{II_b}$ with entropy given by \eqref{segbmax}. In Figure \ref{fig:indphase} we plotted these entropies for $\tilde\zeta$ in the range $0\leq \tilde\zeta\leq \zeta$ for a representative value of $\zeta<\frac{1}{2}$ and another representative value with $\zeta>\frac{1}{2}$. We observe: 
\begin{equation}
    s_{\mathcal I}(\zeta,\tilde\zeta) = \begin{cases}
        s_X(\zeta,\tilde\zeta) &,~~ {\rm if ~~}\tilde\zeta\leq\frac12\\
        s_{max}(\zeta,\tilde\zeta) &,~~ {\rm if ~~} \tilde\zeta> \frac12
    \end{cases}\label{indphases}
\end{equation}
Moreover, the entropy of the two component configuration at point $B$ do not dominate the microcanonical ensemble for any  $\tilde\zeta \in (0,\zeta)$. We see from \eqref{indphases} that the microcanonical index $s_{\mathcal I}$ undergoes a phase transition at $\tilde\zeta = \frac12$: for values of $\tilde\zeta<\frac12$ the index is dominated by the entropy $s_X$ of the pure BPS black hole at point $X$, and for $\tilde\zeta>\frac12$ it is dominated by the entropy $s_{max}$ of the two component phase at the extremum point $q=q_{max}$. The entropy $s_X$ \eqref{indxent} of the pure BPS black hole phase at the phase transition point $\tilde\zeta=\frac12$ is given by  
\begin{equation}
    s_X\left(\zeta,\frac12\right) = \pi \left(-1+\sqrt{2\zeta}\right) ~,
\end{equation}
and the entropy $s_{max}$ \eqref{segbmax} of the two component phase at the same phase transition point given by
\begin{equation}
    s_{max} \left(\zeta,\frac12\right) = \pi \left(-1+\sqrt{2\zeta}\right) ~.
\end{equation}
The agree for all values of $\zeta \geq \frac{1}{2}$. The phase transition at $\tilde\zeta = \frac12$ is of second order because the entropy is continuous across the phase boundary at $\tilde\zeta=\frac12$ but is given by different analytic functions in the two phases.

\begin{figure}
    \centering
    \includegraphics[width=0.5\linewidth]{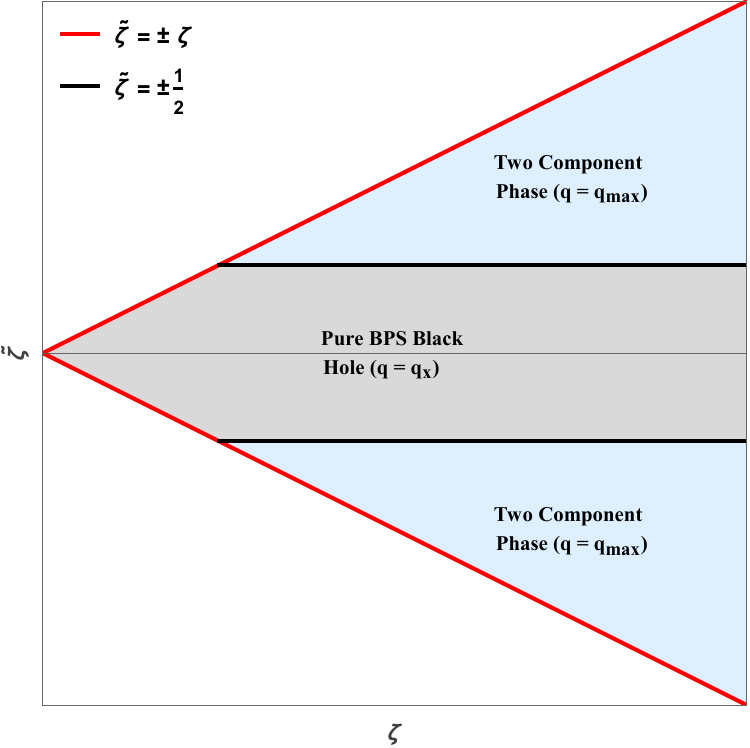}
    \caption{The microcanonical phase diagram for the superconformal index as function of index charges $(\zeta,\tilde\zeta)$ in the physical region $|\tilde\zeta|\leq  \zeta$. The pure BPS black hole phase is shaded grey and the two component phase is shaded blue. There are connected by second order phase transitions at $|\tilde\zeta|=\frac{1}{2}$.}
    \label{fig:indphasediag}
\end{figure}


\subsection{The Phase Transition in the Canonical Ensemble}

The entropy of supersymmetric $AdS_4$ black holes has been accounted for microscopically, by computing the index of the dual CFT$_3$ \cite{Choi:2019zpz}. In the saddlepoint approximation, the results can be summarized by the indicial free energy \cite{Choi:2018fdc} 
\begin{align}
    \mathcal{F}(\Phi_1^\prime,\Phi_2^\prime) = -i k \frac{\Phi_1^\prime \Phi_2^\prime}{\Omega^\prime}  = -i k \frac{\Phi_1^\prime \Phi_2^\prime}{\Phi_1^\prime + \Phi_2^\prime - 2\pi i}
    ~,  \label{entfunc}
\end{align}
where potentials are constrained to satisfy
\begin{equation}
\label{eqn:Phicontr}
    \Phi_1^\prime + \Phi_2^\prime - \Omega^\prime = 2\pi i ~.
\end{equation}
The indicial entropy then follows via the Legendre transform
\begin{align}
    S_{\mathcal I}(\zeta,\tilde\zeta) = {\rm ext}_{\Phi_{1,2}^\prime} \left\{-\mathcal F + k\Phi_1^\prime (\zeta+\tilde\zeta) + k\Phi_2^\prime (\zeta - \tilde\zeta) \right\} ~. \label{indextrem}
\end{align}
It has been found to agree with the entropy of the pure black hole \cite{Cvetic:2005zi,Hristov:2019mqp} with the corresponding charges. 

The extremization in \eqref{indextrem} is carried out several places in the literature, including \cite{Choi:2018fdc,Larsen:2020lhg}. 
For completeness, we work it out in the Appendix \ref{sec:entext}, using our indicial variables $(\zeta, \tilde\zeta)$. We report the values of the complex chemical potentials $\Phi_{1,2}^\prime$ where the extremum is attained in \eqref{appphi1p} and \eqref{appphi2p}). They are
\begin{align}
    \Phi_1^\prime &= i\pi \left(-1 + \frac{1+2i\tilde\zeta}{\sqrt{1-4\tilde\zeta^2+4i\zeta}}\right)~, \\
    \Phi_2^\prime &= i\pi \left(-1 + \frac{1-2i\tilde\zeta}{\sqrt{1-4\tilde\zeta^2+4i\zeta}}\right)~.
\end{align}
It is interesting that, at the phase boundary where $\tilde\zeta = \frac12$, the chemical potential $\Phi_1^\prime$ becomes purely imaginary. The real part ${\rm Re}~\Phi_1'>0$ for 
index lines with $\tilde\zeta<\frac{1}{2}$, but ${\rm Re}~\Phi_1'<0$ for those with $\tilde\zeta>\frac{1}{2}$. In either case $\zeta$ can take any value, as long as $\zeta\geq\tilde\zeta$. For the phase transition at $\tilde\zeta=-\frac{1}{2}$ there are analogous results, with $\Phi_1^\prime$ replaced by $\Phi_2^\prime$.

The comparison between macroscopic black hole entropy and the superconformal index computed in CFT$_3$ is successful also for  
index lines with $|\tilde\zeta|>\frac{1}{2}$. However, for such values of charges, we have found a different configuration that has two components and gives a larger contribution to the same index line. In the canonical ensemble, the new phase at $|\tilde\zeta|>\frac{1}{2}$ is precisely when one of the potentials has negative real part: ${\rm Re}~\Phi_1'<0$ or ${\rm Re}~\Phi_2'<0$. In these cases, the standard extremization of the indicial free energy \eqref{entfunc} cannot be trusted, and there are  new contributions to the index. 


We can understand this result from thermodynamic considerations. We consider $\tilde\zeta>0$ without loss of generality.  
The first law of thermodynamics for BPS black holes is:
\begin{equation}\label{firstlaw}
    \delta s = 2~ {\rm Re}(\Phi_1^\prime) \delta q_{1c} + 2~{\rm Re}(\Phi_2^\prime) \delta q_{2c} + {\rm Re}(\Omega^\prime) \delta j_c ~,
\end{equation}
where $s$ is the BPS entropy \eqref{reg2bentf}, and $(\Phi_{1}^\prime,\Phi_2^\prime,\Omega^\prime)$ are the complex chemical potentials obtained from extremizing the free energy \eqref{entfunc}. Their expressions in terms of the charges $(q_{1c},q_{2c})$ of the BPS black hole are given in \eqref{chempotph12re} and \eqref{chempotomre}. We emphasize that the first law, as written in \eqref{firstlaw}, is a highly non-trivial statement. This is because the primed chemical potentials $(\Phi_{1,2}^\prime, \Omega^\prime)$ are not well defined for BPS black holes \cite{Larsen:2020lhg,Ezroura:2021vrt}. Therefore, in order to formulate the first law in the form of \eqref{firstlaw}, we have used the real parts of the chemical potentials obtained through the extremization procedure described in Appendix \ref{sec:entext}. The first law \eqref{firstlaw} can be validated by substituting the expressions for the chemical potentials, charges  and taking into account that $j_c$ is the function of $(q_{1c},q_{2c})$ given in \eqref{sheetconst}.

The chemical potentials obey the constraint \eqref{eqn:Phicontr}.
In particular, the real parts of the chemical potentials satisfy
\begin{equation}\label{realpotcond}
    {\rm Re}(\Phi_1^\prime) + {\rm Re}(\Phi_2^\prime) - {\rm Re}(\Omega^\prime) = 0 ~.
\end{equation}
Substituting \eqref{realpotcond} back into the first law \eqref{firstlaw} we get
\begin{equation}
\label{eqn:deltas}
    \delta s = {\rm Re}(\Phi_1^\prime)~\left(2\delta q_{1c} + \delta j_c\right) + {\rm Re}(\Phi_2^\prime) ~ (2\delta q_{2c} + \delta j_c) ~.
\end{equation}
On the line segment $i_{II_b}$, the core black holes lie on the plane \eqref{reg2bcondn} where the variations satisfy 
\begin{equation} 
    2\delta q_{2c} + \delta j_c = 0 \label{indplanevar} ~.
\end{equation}
Therefore, it is only the first term in \eqref{eqn:deltas} that contributes. 
Moreover, the core black holes lie on the black hole sheet, so the variations $\delta q_{1c}$ and $\delta j_c$ are not independent. The chain rule
\begin{equation}
     \delta j_c = \frac{\partial j_c}{\partial q_{1c}} \delta q_{1c} + \frac{\partial j_c}{\partial q_{2c}} \delta q_{2c} ~,
\end{equation}
and the condition \eqref{indplanevar} gives 
\begin{equation}
    \delta s = 2~ {\rm Re}(\Phi_1^\prime)~\left(1+ \frac{\frac{\partial j_c}{\partial q_{1c}}}{2+\frac{\partial j_c}{\partial q_{2c}}} \right) \delta q_{1c} 
    = 2~ {\rm Re}(\Phi_1^\prime)~\left(\frac{8 \left(q_{1c}^2+6 q_{1c} q_{2c}+q_{2c}^2\right)+2}{8 q_{1c} (q_{1c}+3 q_{2c})+\sqrt{16 q_{1c} q_{2c}+1}+1} \right) \delta q_{1c}~.
\end{equation}
In the second expression we used \eqref{sheetconst} for $j_c$ and notice that it is positive for all $q_{1c},~q_{2c}>0$. We conclude that, if the entropy is maximized along the line segment $i_{II_b}$, it happens precisely at the point where ${\rm Re}~\Phi_1^\prime~= 0$. 

Our result should not really be surprising. The superconformal index \eqref{indprel} is illdefined when ${\rm Re} ~\Phi_{1,2}^\prime<0$. Therefore, when extremization of the indicial free energy \eqref{entfunc} leads to ${\rm Re}~ \Phi_{1,2}^\prime<0$, some aspect of the prescription must be modified. Our two component configurations resolve the tension, because they involve core black holes with ${\rm Re} ~\Phi_{1}^\prime=0$ or ${\rm Re}~ \Phi_{2}^\prime=0$ in the entire region where the candidate pure BPS black hole would give a potential with negative real part. 



In summary, in this section we have identified phase transitions in the index. In the microcanonical ensemble, they are when indicial charges $|\tilde\zeta|=\frac{1}{2}$ and, in the canonical ensemble, they are when one of ${\rm Re} ~\Phi_{1,2}^\prime=0$. In the next section we will study the complex BPS black holes in more detail and try to identify instabilities in these solutions beyond the phase boundary.

\section{Allowable Complex BPS Black Holes}
\label{sec:allowable}
In this section we develop the analytical continuation of the AdS$_4$ black holes that preserve supersymmetry. 
We determine the regions of parameter space where the complex black hole permits an interpretation in statistical physics and study the KSW conditions on allowable Euclidean geometries. 

\subsection{Analytical Continuation to Complex Variables}
\label{sec:ancont}

In our presentation of black hole solutions in Section \ref{sec:2chargesol} all variables were presumed real. In that context saturation of the BPS bound on the mass \eqref{eqn:BPSineq} imposes two conditions on the four conserved charges $(M, J, Q_1, Q_2)$, rather than one. Equivalently, it imposes two conditions on the four parameters $(m, a, \delta_1, \delta_2)$ that give all physical variables. This situation follows from the equation $\Delta_r(r_+)=0$ that gives the coordinate position $r_+$ of the event horizon. The explicit expression \eqref{eqn:DelrBPS} for the conformal factor $\Delta_r$ sets the sum of two complete squares to zero. When all variables are real, each term must vanish by itself. This gives two conditions on the real variables.  

We can circumvent this conclusion by relaxing the reality assumption.  The new possibility that arises because some variables are complex is that the horizon equation $\Delta_r(r_+)=0$ given in \eqref{eqn:DelrBPS} can be solved by imposing a single condition. We take: 
\begin{equation}
q = (1+ag) r_+ \pm  i g \big( r^2_+ - r^2_*\big)~.
\label{eqn:qcomplex}
\end{equation}
This solution is not unique, but it gives a simple construction where the analytical continuation to the complex plane involves a single real parameter. 

To specify the complex solution completely, we also impose BPS saturation. This amounts to the equality \eqref{eqn:BPScon} which we repeat here:  
\begin{equation}
    1 + ag=\coth(\delta_1+\delta_2)~.
    \label{eqn:BPScon2} 
\end{equation}
This condition applies also when the parameters are complex. 
Once parameters are required to satisfy \eqref{eqn:qcomplex} and \eqref{eqn:BPScon2}, formulae in Section \ref{sec:2chargesol} give all physical variables of interest as function of three real parameters $(a, r_+, r_*)$. Specifically, $m$ follows by combining \eqref{eqn:qcomplex} with the definition of $q$ \eqref{eqn:qdef}. The definition of $r_*$ \eqref{eqn:rstardef} gives $m^2(s^2_1-s^2_2)^2=r^2_*-ag^{-1}$. Taken together, we have $(m, q, \delta_1, \delta_2)$ as functions of $(a, r_+, r_*)$. This in turn gives 
the physical variables in microcanonical ensemble (\ref{edef}-\ref{q2def}) and canonical ensemble (\ref{eqn:betapar}-\ref{eqn:phi2par}). 

It is important to be explicit about what is complex, and what is not. 
We keep $a$ real so, because of the BPS condition \eqref{eqn:BPScon2}, $\delta_1 + \delta_2$ must also be real. Then the definition \eqref{eqn:qdef} of $q$ shows that $m$ and $q$ have the same phase, the one that follows from \eqref{eqn:qcomplex}. We also keep $r_*$ real so the definition of $r_*$ \eqref{eqn:rstardef} requires that\footnote{We have not considered the branch where $m(s^2_1-s^2_2)$ is purely imaginary. More generally, we do not claim that the analytical continuation we specify in detail is the most general. However, it appears to be the simplest. }
$$
m(s^2_1-s^2_2) = m \tanh(\delta_1-\delta_2)\sinh(\delta_1+\delta_2)~,
$$
remains real. Therefore, $\coth(\delta_1-\delta_2)$ acquires the same phase as $m$ and $q$. Altogether, the parameters $(m, q, \delta_1, \delta_2)$ only involve a single complex phase.  

The analytical continuation makes the conformal factor $\Delta_r^*(r)$ \eqref{eqn:DelrBPS} complex. It can be presented concisely as
\begin{eqnarray}
\Delta_r^* (r) &=& \Big( -(1+ag)(r-r_+)\pm i g(r_+^2-r^2_*)\Big)^2 + g^2(r^2 - r^2_*)^2~.
\label{eqn:DerBPS}
\end{eqnarray}
The equation $\Delta_r^* (r)=0$ has the solution $r=r_+$, as it should. Later in this section we show that all other aspects of the metric remain real, also after analytical continuation of the black hole parameters. In the limit $r_+\to r_*$, the conformal factor $\Delta_r^*$ given in \eqref{eqn:DerBPS} becomes real we well, so then the entire geometry is real. In this section we refer to the surface $r_+=r_*$ in parameter space as the {\it real} BPS surface, because ``real" stresses that the geometries are not complex. On the real BPS surface, the complex solutions reduce to the BPS black holes on the black hole sheet. 

Even though $\Delta_r^*(r)=0$ is a quartic equation in $r$, its four roots can be found in closed form. 
Taking the upper sign in \eqref{eqn:DerBPS}, they are
\begin{align}
r  = \begin{cases}
r_+\\
 i g^{-1}(1+ag) - r_+ \\
 -\frac{1}{2g} i (1+ag) \pm \frac{1}{2g}\sqrt{ 8g^2r^2_* - 4g^2r^2_+ - (1+ag)^2 + 4i(1+ag)gr_+}
 \end{cases}
\end{align}
The complex roots do not form complex conjugate pairs, because the quartic \eqref{eqn:DerBPS} has complex coefficients. 
The four roots of the equation $\Delta_r^*(r)=0$ with lower sign are the complex conjugates of the four roots given above.
In the real BPS limit $r_+\to r_*$, the two roots in the last line reduce to $r_+$ and $-ig^{-1}(1+ag)-r_+$, respectively. Thus $r_+$ becomes a double root and $\pm ig^{-1} (1+ag) - r_+$ becomes a pair of complex conjugate roots. If we want to designate an inner horizon position $r=r_-$, the best option may be the root involving $+\sqrt{\cdots}$, because that is the root that approaches $r_+$ on the real BPS surface. 

\subsection{Complex Potentials and Charges}
In the previous subsection we specified the complex solutions in terms of the three real parameters $(a, r_+, r_*)$, with other parameters specified  
by the conditions \eqref{eqn:qcomplex} and \eqref{eqn:BPScon2}. We now use these values to compute physical potentials and charges. 

For a general black hole, the inverse temperature $\beta$ is given by \eqref{eqn:betapar}. The complex temperature then follows, by differentiation of the conformal factor given in \eqref{eqn:DerBPS}: 
\begin{eqnarray}
T & = & \left.\frac{\partial_r \Delta_r^*}{4\pi (r_1 r_2 + a^2)} \right|_{r=r_+} =  \frac{2g^2r_+ \mp  i g (1+ag)}{2\pi}  \frac{r^2_+ - r^2_*}{r^2_+ - r^2_* + ag^{-1}(1+ag)}~.
\label{eqn:compT}
\end{eqnarray}
The limit of vanishing temperature $T\to 0$ coincides with the limit $r_+\to r_*$ where the geometries become real. 

The rotational velocity is given in \eqref{eqn:omegapar}. The corresponding BPS potential conjugate to angular momentum is 
\begin{eqnarray}
 \Omega' =\beta( g-\Omega) & = & \frac{2\pi(1-ag)}{2gr_+ \mp i(1+ag)}~. 
  \label{eqn:Omegaprime}
\end{eqnarray}
Finally, the electric potentials at the horizon are (\ref{eqn:phi1par}-\ref{eqn:phi2par}), so the BPS potential 
for total charge gives: 
\begin{eqnarray}
\Phi_1' + \Phi_2'  & = &  g\beta(2-\Phi_1 - \Phi_2)=  \frac{4\pi g(-a\mp i r_+)}{2gr_+ \mp i(1+ag)}~.
 \label{eqn:Phiprimesum}
\end{eqnarray}
These potentials satisfy the sum rule required by supersymmetric boundary conditions: 
\begin{eqnarray}
 \Phi_1' + \Phi_2' - \Omega' 
  = \mp 2\pi i~.
 \label{eqn:potentialconstr}
 \end{eqnarray}
The computations were done {\it without} imposing any condition on the temperature, so this sum rule applies to entire space of solutions, dependent on the three real parameters $(a,r_+,r_*)$.

In order to access the individual electric potentials, rather than just the sum, we introduce: 
\begin{eqnarray}
 \Phi_1'- \Phi_2'     & = &  -g\beta(\Phi_1 - \Phi_2)= \frac{\pm 4\pi i  m g (s_1^2 - s^2_2) }{2gr_+ \mp i (1+ag)}=  \frac{\pm 4\pi ig  \sqrt{r^2_*-ag^{-1}}}{2gr_+ \mp i (1+ag)}
~.
 \label{eqn:Phiprimediff}
\end{eqnarray}
In the second equality we use $m(s^2_1 - s^2_2)= \sqrt{r^2_* - ag^{-1}}$. That is slightly imprecise. The equations to the left and in the middle are odd under $1\leftrightarrow 2$. In the final equation this sign is absorbed into the overall $\pm$. 

Because of the constraint \eqref{eqn:potentialconstr}, we can interpret the sum $\Phi_1' + \Phi_2'$ as a dependent variable. That leaves two complex variables $\Omega' $ and $\Phi_1' - \Phi_2'$. However, comparing the explicit formulae \eqref{eqn:Omegaprime} and \eqref{eqn:Phiprimediff}, we see that the phases of these complex potentials are related: they differ by exactly $\frac{\pi}{2}$. Therefore, the three complex potentials $\Omega' , \Phi_1', \Phi_2'$ parametrize a space with three real dimensions, as they should. 

Interestingly, $\Omega\ell-1$ is entirely real and $\Phi_1 - \Phi_2$ is purely imaginary for all the complex BPS solutions. Therefore, the nontrivial phase carried by the BPS potentials $\Omega' $ and $\Phi_1' - \Phi_2'$ is inherited from the complex temperature \eqref{eqn:compT}. This observation applies also on the real BPS surface where $\Omega\ell-1$, $\Phi_{1,2}-1$, and $T$ all approach zero at the same rate, such that their ratios $\Omega', \Phi'_{1,2}$ become finite complex numbers. In this limit the absolute value $|T|\to 0$, but the phase of $T$ coincides with the phase of other variables. 

For the complex BPS solutions, the black hole entropy \eqref{eqn:BHentropy} becomes
$$
S = \frac{\pi}{G_4} \left( \frac{r^2_+- r^2_*}{1-a^2g^2} + \frac{ag^{-1}}{1-ag}\right)~. 
$$
It is always real.\footnote{This is a special feature for AdS$_4$ black holes. For example, the analogous entropies for AdS$_3$ \cite{Larsen:2021wnu} and AdS$_5$ \cite{Cabo-Bizet:2018ehj} black holes both have an imaginary part.} The corresponding on-shell action
\begin{eqnarray}
I & =&  - S + \beta ( M - \Omega J - 2\Phi_1 Q_1 - 2\Phi_2 Q_2)
\cr
& = & - S + \beta ( M- J - 2Q_1 - 2Q_2) + (\Omega' J + 2\Phi'_1 Q_1\ell + 2\Phi'_2 Q_2\ell)~, 
\end{eqnarray}
can be evaluated with the result
\begin{equation}
I = \frac{\pi  \left(a^2 g+a (-1\pm2 i g r_+)+g \left(r_*^2-r_+^2\right)\right)}{g ~G (a g-1) (a g\pm2 i g r_++1)}
= \pm ik\frac{\Phi'_1\Phi'_2}{\Omega'} ~.
\end{equation}
This is the generalization of the HHZ-potential \cite{Hosseini:2017mds} to the AdS$_4$ setting \cite{Choi:2018fdc}. It applies to the full family of complex BPS solutions, depending on three real parameters. The computation did not specialize to $r_+=r_*$ at any point. It is a derivation of \eqref{entfunc} from a gravitational point of view. 

The conserved charges $(J, Q_1,Q_2)$ that govern the microcanonical ensemble are also complex. They are related to 
the three real parameters $(a, r_+, r_*)$ through:  
\begin{eqnarray}
J & =&  \frac{ar_+}{G(1-ag)^2}\Big( 1 \pm i g\frac{r^2_+ - r^2_*}{(1+ag)r_+}\Big)~,\cr
Q_1 + Q_2  & =&   \frac{r_+}{2G(1-ag)}\Big( 1 \pm ig \frac{r^2_+ - r^2_*}{(1+ag)r_+}\Big)~,\cr
Q_1 - Q_2 & =& {\rm sgn}(Q_1-Q_2) \cdot\frac{\sqrt{r^2_*-ag^{-1}}}{2G(1-ag)} ~.
\label{eqn:physJQs}
\end{eqnarray}
We present the combinations $Q_1\pm Q_2$, rather than $Q_1$ and $Q_2$ by themselves, because this form highlights that 
$J$ and $Q_1 + Q_2$ have the same phase, while $Q_1-Q_2$ is real. The complex $J$ and $Q_1 + Q_2$ represent three real parameters, because they have the same phase. They parametrize all the complex BPS solutions, because the remaining conserved charge is a dependent variable: 
\begin{equation}
| Q_1 - Q_2 | = \sqrt{ g~{\rm Re} (Q_1 + Q_2) - g\frac{1+ag}{2(1-ag)} {\rm Im} (Q_1 + Q_2) - \frac{ag^{3}}{4(1-ag)^2}}~. 
\end{equation}
Here
\begin{equation}
a = \frac{J}{M} = \frac{J}{ Jg + 2(Q_1 + Q_2)}~, 
\end{equation}
is a real variable even when the charges are complex, because $J$ and $Q_1 + Q_2$ have the same phase. 
On the real BPS surface $r_+=r_*$ the charges $(J, Q_1, Q_2)$ all become real. 
This is in contrast to the conjugate potentials
(\ref{eqn:compT},
\ref{eqn:Omegaprime}, 
\ref{eqn:Phiprimesum},
\ref{eqn:Phiprimediff}) which retain a phase on the real BPS surface.

In summary, the BPS black holes depend on three real parameters. Generally, the geometries are complex, but a two-parameter subfamily is real. This is reminiscent of the BPS phase diagram which depends on three real parameters, taken to be the conserved charges $(J, Q_1, Q_2)$. In that context the gravitational configuration generally has two distinct components, but a two-parameter subfamily is a pure BPS black holes. 
It is the same two-parameter family of BPS black holes that appear in these distinct settings. However, the continuations to a three parameter family of complex solutions 
and a three parameter family of two component configurations do not have any obvious relation. 

\subsection{Physical Conditions on Complex Potentials}
In this subsection we consider the physical conditions that must be imposed on the complex BPS solutions with focus on the potentials. This is related to the discussion of physical conditions on the index in Section \ref{sec:supindex}. However, that discussion was based on real solutions, including two component configurations. Here we consider only pure BPS black holes, but we allow complex solutions. 

{\it Causality in the boundary theory} requires $|\Omega|<1$.
Since $\Omega$ given by \eqref{eqn:omegapar} is real and positive this amounts to: 
$$
\Omega  = \frac{a(1 + g^2r_{1+} r_{2+})}{r_{1+} r_{2+} + a^2} < g ~~\Leftrightarrow~~ag^{-1} < r_{1+} r_{2+}~.
$$
The shifted coordinates \eqref{eqn:shifteddef} and the definition \eqref{eqn:rstardef} combine to $r_{1+} r_{2+} = r^2_+ - r^2_* + ag^{-1}$, so the condition $|\Omega|<1$ becomes: 
\begin{equation}
r_* <  r_+~.
\label{eqn:Omegacond}
\end{equation}
Geometrically, the deformation away from the real BPS surface must be such that the coordinate position $r_+$ has increased, rather than decreased. We interpret $r^2_+-r^2_*$ as a regulator needed to define the real BPS surface precisely, so we exclude the strict limit $r_+ = r_*$. 

Additional physical conditions arise from the presumption that supersymmetric black hole solutions permit an interpretation as {\it a statistical ensemble of quantum states} with index:  
\begin{eqnarray}
I & = & {\rm Tr}\left[  (-)^F e^{ - \beta(E - \Omega J - 2\Phi_1 Q_1 - 2\Phi_2 Q_2)} \right]\cr
& = & {\rm Tr}\left[  (-)^F e^{ - \beta\{Q , Q^\dagger \}  - \Omega' J -  2\Phi'_1 Q_1 - 2\Phi'_2 Q_2} \right]~.
\label{eqn:indexdef}
\end{eqnarray}
The superalgebra is written schematically as the anticommutator $\{Q , Q^\dagger \} = E - J - 2Q_1 - 2Q_2$. It is nonnegative on all physical states and vanishes precisely on the BPS states. 
The index is independent of the inverse temperature $\beta$. However, $\beta$ serves as a regulator, so it must ensure that 
the index is manifestly convergent. Therefore, we require ${\rm Re}~\beta>0$, the real part of the complex temperature  \eqref{eqn:compT} must be positive. This gives another path to the condition $r_*<r_+$ given in \eqref{eqn:Omegacond}.

The primed potentials $\Omega' = \beta(g-\Omega)$, $\Phi'_{1,2} = g\beta(1-\Phi_{1,2})$ can be complex but, to ensure convergence, their real parts must be 
positive ${\rm Re}~\Omega'>0$, ${\rm Re}~\Phi'_{1,2}>0$.\footnote{In the limit $T=\beta^{-1}\to 0$, the prime reduces to the derivative with respect to temperature, {\it except for the sign}. With the convention here, positive real potentials correspond to suppression, following the standard in statistical mechanics.} The physical condition that ${\rm Re}~\Omega'>0$ amounts to the requirement that $ag<1$, according to \eqref{eqn:Omegaprime}. This is rather trivial, because geometries with $ag\geq 1$ have
$\Delta_\theta=0$ for some real values of the polar angle $\theta$, and so they are manifestly singular.

The analogous conditions on the complex electric potentials  (\ref{eqn:potentialconstr}-\ref{eqn:Phiprimediff}) are more interesting. The potential
$$
\Phi'_1 = - 2\pi g \frac{a \pm i r_+ \mp i m (s^2_1 - s^2_2)}{2gr_+ \mp i (1+ag)}~,
$$
has real part
$$
{\rm Re}~ \Phi'_1 = 2\pi g \frac{(1-ag) r_+  - (1+ag)m  (s^2_1 - s^2_2)}{(2gr_+)^2 +  (1+ag)^2}~.
$$
The analogous formulae for $\Phi'_2$ follow by the interchange $1\leftrightarrow 2$. 
Recalling that $m  |s^2_1 - s^2_2|  = \sqrt{r^2_*-ag^{-1}}$, according to the definition \eqref{eqn:rstardef}, the condition ${\rm Re}~ \Phi'_k>0$ is violated for either 
$k=1$ or $k=2$ when 
\begin{equation}
(1-ag) r_+ <  (1+ag)  \sqrt{r^2_*-ag^{-1}}~.
\label{eqn:chargeasym}
\end{equation} 
Both sides of this inequality are real and positive. 
Trading the parameters $(a, r_+, r_*)$ for physical charges via \eqref{eqn:physJQs}, the unstable region becomes: 
\begin{equation}\label{compchargeconst}
\frac{Q_1 + Q_2}{gJ+Q_1 + Q_2} < \frac{|Q_1-Q_2|}{{\rm Re} (Q_1 + Q_2)}~.
\end{equation}
Thus the inequality \eqref{eqn:chargeasym} expresses an instability when the two electric charges become too unbalanced.

The physical significance of general complex BPS geometries is unclear, as discussed in the final paragraph of the preceding subsection, but the situation is clear in the limit $r_+ \to r_*$ where the geometry becomes real. We can rewrite the unstable region \eqref{eqn:chargeasym} as: 
\begin{equation}
(1+ag)^2 < 4g^2r^2_+ 
~.
\label{eqn:stabineq}
\end{equation}
In the microcanonical ensemble, the charges \eqref{eqn:physJQs} on the real BPS surface $r_+=r_*$, give 
\begin{equation}
    k^{-1}|Q_1 \ell - Q_2 \ell| > \frac12~. 
\end{equation}
This is precisely the location of the phase transition \eqref{indmaxcond} derived in section \ref{sec:maxsegb}. The reasoning there is entirely Lorentzian and invokes the two component picture, while here we appeal to the complex potentials of a pure BPS black hole. The quantitative agreement follows because the real part of the complex potentials is identical to the physical potential of the core black hole. 

\begin{figure}
    \centering
    \includegraphics[width=0.6\linewidth]{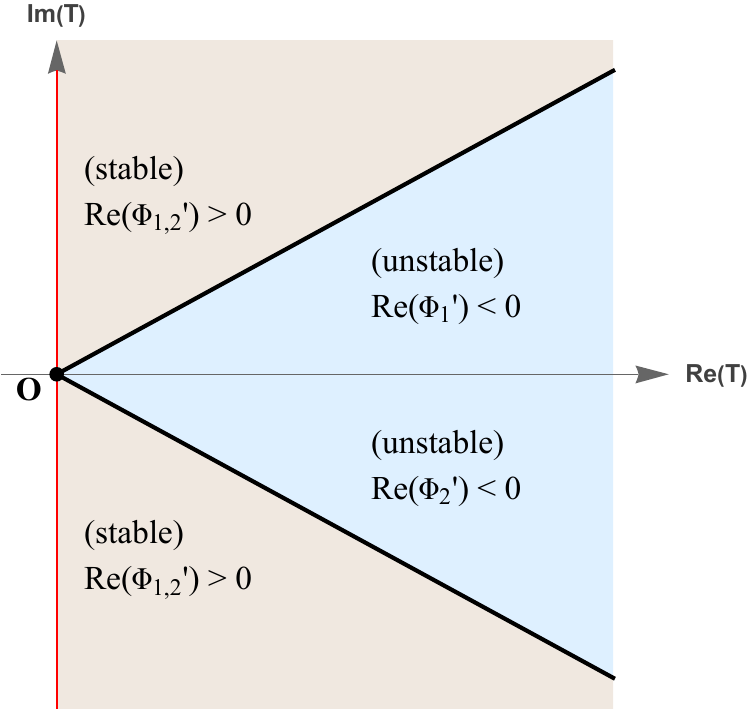}
    \caption{Stability of complex BPS solutions near the real BPS surface represented as phases in the complex $T$ plane. The physical region $\mathrm{Re}\beta > 0$ is to the right of the (red) $y$-axis. The strict limit of real BPS geometry is at the origin $T=0$, but the phase ${\rm Arg} T$ has physical significance. In the limit $|T| \rightarrow 0$, the solutions inside the (blue) wedge defined by $|{\rm Arg}(T)| < \frac{\pi}{4}$ are unstable, since either $\mathrm{Re}\Phi_1'<0$ or $\mathrm{Re}\Phi_2'<0$ in this region.}
    \label{fig:compstability}
\end{figure}

In the canonical ensemble, we can interpret the condition \eqref{eqn:stabineq} as a remnant of the complex phase that persists on the real BPS surface. For example, the modulus of the temperature \eqref{eqn:compT} $|T|\to 0$, but its phase has a finite limit: ${\rm Arg}~ T = {\rm Arg}~  ( 2gr_+  \mp i (1+ag))$. Therefore, the unstable region \eqref{eqn:stabineq} corresponds to 
$|{\rm Im}~T| <|{\rm Re}~T|$. 
Combining with the condition ${\rm Re}~ T>0$ discussed after \eqref{eqn:indexdef}, the stable region becomes 
\begin{equation}
\frac{\pi}{4} <  | {\rm Arg} T | <\frac{\pi}{2} ~.
\label{eqn:argTrange}
\end{equation}
This form of the stability criteria is plotted in Figure \ref{fig:compstability}. At some level, it is awkward to impart physical significance to the phase of $T$ in the limit where $|T|\to 0$. On the other hand, it is equivalent to several finite measures. For example, the potential conjugate to angular momentum $\Omega' = \beta(g-\Omega)$ has the same phase (except for a sign), because $\Omega$ is real on the real BPS surface, and the electric potentials $\Phi'_{1,2}$ similarly inherit their phases from $T$. The phase of the temperature $T$ does not give preference to any of the charges. 

\subsection{Euclidean Geometries and the Partition Function}
Up to this point, the complex BPS solutions discussed in this section were discussed entirely in terms of thermodynamic variables and presented in Lorentzian signature. That is convenient, because it does not require any new notations, but the physical content is predicated on an analytical continuation to Euclidean signature that can be interpreted as a saddle of the gravitational path integral. That is far from trivial. 

Our main focus in this section as a whole are the complex BPS geometries but, in this subsection, we consider the continuation to Euclidean signature of
all real Lorentzian black holes, without imposing supersymmetry. This is simpler, and will turn out to be an important ingredient in the study, presented in the next subsection, of the complex geometries singled out by supersymmetry. In the context of the gravitational path integral, this subsection discusses the Euclidean geometries that contribute to the partition function, while the next subsection considers the analogous question for the index. 

The black hole solutions we consider are somewhat elaborate, but they nevertheless enjoy several simplifications.
They have distinct timelike and spacelike Killing vectors: they are independent of time ``$t$" and azimuthal angle ``$\phi$". The remaining coordinates can be chosen as a``radial" coordinate $r$ and a ``polar" angle $\theta$ that do not mix with $t, \phi$, or one another. Therefore, the metric can be presented in the form:\footnote{Large classes of stationary supergravity black holes that preserve at least two supersymmetries have analogous simplifications, whatever the dimension.} 
\begin{equation}
ds^2_4 = g_{ij} dx^i dx^j = g_{\mu\nu} dx^\mu dx^\nu  + g_{\theta\theta} d\theta^2 + g_{rr} dr^2 ~. 
\label{eqn:defmetric}
\end{equation}
The latin indices can take all four values while the greek letters $\mu, \nu$ denote $t, \phi$. All nontrivial metric components may depend on $r$ and $\theta$, but not on $t, \phi$.\footnote{As presented here, the simplifying properties depend on the choices of coordinates, but they can be recast in a covariant form. A metric that can be presented in the form \eqref{eqn:defmetric} is orthogonally transitive with respect to the two commuting Killing vectors. For relevant discussion and entry points to the literature, see \cite{Keeler:2012mq}}

According to the proposal by KSW, the allowable Euclidean metrics are those that satisfy: 
\begin{equation}
{\rm Re} \left( \sqrt{g} g^{i_1 j_1} \cdots g^{i_p j_p} \right)F_{i_1\cdots i_p}F_{j_1\cdots j_p}\geq 0~,
\label{eqn:KSWmet}
\end{equation}
for all real p-forms $F$. In the examples we study it is sufficient to consider the KSW condition for $p=1$ with indices in the directions singled out by Killing symmetries: 
\begin{equation}
{\rm Re} \left( \sqrt{g} g^{\mu\nu} \right)F_{\mu}F_{\nu}\geq 0~.
\label{eqn:KSWmet2}
\end{equation}
For a real Lorentzian metric of the form \eqref{eqn:defmetric}, the Euclidean continuation $t\to - i t_E$ takes $g_{tt}\to - g_{tt}$ and $g_{t\phi}\to - i g_{t\phi}$. The determinant of the metric remains real and $g_{t\phi}$ drops out, because the condition \eqref{eqn:KSWmet2} explicitly takes the real part. 
Therefore, in this case, the KSW conditions amount to 
\begin{eqnarray}
g_{\phi\phi} &>& 0~,
\cr
g_{tt} &<& 0~. 
\label{eqn:KSWbasic1}
\end{eqnarray}
The notation refers to the original Lorentzian metric \eqref{eqn:defmetric} in the region outside the black hole event horizon, before continuation to Euclidean signature. Geometrically, they demand the existence of a global ``time" $t$, which can then be continued to Euclidean signature $t\to - i t_E$ with no obstruction.

In the following, we consider the full four-parameter family of two-charge AdS$_4$ black holes with real parameters, with no condition of supersymmetry, nor on extremality. 
The geometry of these black holes is often presented in Boyer-Lindquist coordinates, as in \eqref{eqn:Lormetric}. In these coordinates, the leading behavior at infinity takes the form $r^2 \Big( ... dt^2 + ...(d\phi_0 + ag^2dt)^2\Big)$. Therefore, the $S^2$ at infinity rotates with angular velocity $\partial_t\phi_0 = -ag^2$. Motivated by the dual CFT$_3$, it is common to address this by introducing $\phi_1 = \phi_0 + a g^2 t$, because that gives coordinates with the canonical asymptotic behavior where the $S^2$ at infinity is static.  

However, in such coordinates the geometry has near horizon structure: 
$$
ds^2_4 = - ...\Delta_r d\phi_1^2  + ...(dt  - \Omega^{-1} d\phi_1)^2~. 
$$
Factors indicated schematically by dots are positive outside the black hole event horizon and the conformal factor $\Delta_r>0$ as well. 
In the near horizon region, it is $\phi_1$ that is time-like, so a worldline violates causality, if it is parametrized by $t$ and has fixed $\phi_1$. In other words, there is an ergoregion.

This motivates yet another angular coordinate $\phi = \phi_1 - \Omega t =  \phi_0 - (\Omega -ag^2)t $. In these coordinates, the metric \eqref{eqn:Lormetric} becomes: 
\begin{eqnarray}
ds^2_4 & = & - \frac{\Delta_r}{\rho^2}  \left( \frac{r_{1+}r_{2+}+a^2\cos^2\theta}{r_{1+} r_{2+} + a^2} dt   - \frac{a\sin^2\theta}{1-a^2g^2} d\phi\right)^2
+ \frac{\rho^2}{\Delta_r}dr^2 +  \frac{\rho^2}{\Delta_\theta}d\theta^2 \cr
&&+  \frac{\Delta_\theta\sin^2\theta}{\rho^2} 
\left( \frac{a(r_+^2 - r^2)}{r_{1+} r_{2+} + a^2}dt -  \frac{(r_1 r_2+a^2)}{1-a^2g^2} d\phi\right)^2 
~,
\label{eqn:explicitmet}
\end{eqnarray}
where 
\begin{eqnarray}
\Delta_\theta & = & 1 - a^2g^2\cos^2\theta ~,\cr
\rho^2 & = & r_1 r_2 + a^2 \cos^2\theta ~.
\label{eqn:delthetadef}
\end{eqnarray}
The variables evaluated at the horizon, such as $r_{1+} r_{2+}$, arise because the rotational velocity $\Omega$ \eqref{eqn:omegapar} is given by an expression that is evaluated at the horizon. 

The coordinates employed in \eqref{eqn:explicitmet} were designed so that the coordinate $t$ is globally defined for any fixed $\phi$. Therefore, in these coordinates, we expect that the standard Euclidean continuation $t\to -i t_E$ is well-defined. In the remainder of this subsection we verify this expectation by explicit computation. 
 
For the first condition in \eqref{eqn:KSWbasic1}, we begin by showing the inequality: 
\begin{eqnarray}
 (r_1 r_2+a^2)(g^2r_1 r_2+1) - \Delta_r & = & 2mr' - r^{\prime 2} + r_1 r_2 \cr
& = & 2mr' (1 + s^2_1 + s^2_2) + 4m^2s^2_1 s^2_2>0~.
\label{eqn:gphilemma}
\end{eqnarray}
We inserted \eqref{eqn:deltar} for $\Delta_r$ and then \eqref{eqn:shifteddef} for $r_{1,2}$. 
When combining the inequality \eqref{eqn:gphilemma} with the formula for $g_{\phi\phi}$ in \eqref{eqn:explicitmet}, we find $g_{\phi\phi}>0$. That is what we wanted to show.

For the second condition in \eqref{eqn:KSWbasic1}, it is sufficient to show
\begin{eqnarray}
(r_{1+}r_{2+}+a^2\cos^2\theta)^2 &>& \Delta_\theta a^2\sin^2\theta~,\cr
\Delta_r  &>&  (r^2-r^2_+)^2 ~.
\label{eqn:KSW2conds}
\end{eqnarray}
The first of these gives
$$
r^2_{1+}r^2_{2+}+ 2a^2r_{1+}r_{2+} \cos^2\theta  + a^4\cos^2\theta- a^2\sin^2\theta>0~.
$$
This is satisfied for all $r_{1+}r_{2+} >a$, which is equivalent to
$$
r^2_+  \geq a + m^2 (s^2_1 - s^2_2)^2 = r^2_*~.
$$
This is the condition \eqref{eqn:Omegacond} that was already imposed for several reasons, including positivity of the temperature ${\rm Re}T>0$. 

To show the second inequality in \eqref{eqn:KSW2conds}, we write the conformal factor \eqref{eqn:deltar} in terms of the shifted coordinate $r$ defined through \eqref{rdef} as 
\begin{equation}
    \Delta_r = g^2r^4 + r^2((1 + ag)^2-  2g^2r^2_*)-2q(1+ag) r +q^2+g^2r_*^4 
\end{equation}
where the parameters $q$ and $r_*$ are related to the other parameters via \eqref{eqn:qdef} and \eqref{eqn:rstardef} respectively. We rewrite the $\Delta_r$ expression given above as
\begin{equation}
\Delta_r = g^2(r-r_+)(r-r_-) u(r) ~, 
\label{eqn:delrfactor}
\end{equation}
where $r_+, r_-$ are related to other parameters through
\begin{eqnarray}
2q(1+ag) & = & g^2(r_++r_-)\left(\frac{q^2+g^2r_*^4}{g^2r_+r_-}-r_+r_-\right)~,\cr
(1 + ag)^2-  2g^2r^2_* & = & \frac{q^2 +g^2r^4_*}{r_+r_-} + g^2r_+ r_- - g^2(r_+ + r_-)^2~, 
\end{eqnarray}
and the remainder polynomial is
\begin{equation}
u(r) = r^2 + r(r_++r_-) + \frac{q^2 +g^2r_*^4}{g^2r_+r_-}~. 
\end{equation}
Inserting the estimate 
\begin{eqnarray}
u(r) &  = &  (r+r_+)(r+r_-) +\frac{2q(1+ag)}{g^2(r_++r_-)}\cr
&>& (r+r_+)(r+r_-)~,
\end{eqnarray}
in \eqref{eqn:delrfactor}, we validate the second inequality in \eqref{eqn:KSW2conds}. That is what we wanted to show. 

\subsection{Euclidean Geometries and the Index}
In the previous subsection we verified the KSW condition \eqref{eqn:KSWmet} for the obvious Euclidean continuation of the Lorentzian black holes with 
metric \eqref{eqn:explicitmet}. This shows that the Euclidean black holes contribute to the Euclidean integral that computes partition function. The computation of the index adds two complications.   First, the analytical continuation of the coordinates generate complex periodicity of the Euclidean time. 
Second, as discussed in subsection \ref{sec:ancont}, the parameters that appear in the Lorentzian geometry may be complex from the outset, to preserve supersymmetry.\footnote{Complex parameters in the geometry appeared already in the original work on the gravitational path integral for Euclidean black holes by Gibbons and Hawking \cite{Gibbons:1976ue}. They assumed that all allowable Euclidean geometries are real so, for Kerr black holes in four asymptotically flat dimensions, they analytically continued the parameter $a=\frac{J}{M}$ as $a\to ia$. We preserve supersymmetry be complexifying some parameters, but we keep $a$ real.}

To make the complex periodicity explicit, we define the analytical continuation as $t\to - i \beta t_E$. This will exhibit the complex temperature $T=\beta^{-1}$ clearly, because the Euclidean time $t_E$ has canonical periodicity $2\pi$. This is significant because the determinant of the Lorentzian geometry \eqref{eqn:explicitmet} is remarkably simple:\footnote{This is not at all obvious, but it is true, also for the complex metrics that compute the index. The analogous simplification also applies to other large classes of black hole solutions in ${\cal N}\geq 2$ supergravity}
\begin{equation}
\sqrt{- {\rm det} g} = \frac{\rho^2 \sin\theta}{1-a^2}~. 
\end{equation}
In particular, it is {\it real}. Therefore, analytical continuation $\sqrt{- {\rm det} g} \to \beta\sqrt{{\rm det} g}$ just gives an overall factor 
and the KSW condition \eqref{eqn:KSWmet2} takes the form: 
\begin{equation}
{\rm Re} \begin{pmatrix} -\beta g_{tt} & -i g_{t\phi} & 0 & 0 \cr -i  g_{t\phi} &\beta^{-1}g_{\phi\phi}& 0 & 0 
\cr  0 & 0 & \beta^{-1}g_{rr} & 0
\cr  0 & 0 & 0 & \beta^{-1}g_{\theta\theta} 
 \end{pmatrix} >0~. 
\label{eqn:KSWBPS2}
\end{equation}
Even though we consider Euclidean black holes, we have not changed the notation for the metric: the components $g_{ij}$ in this formula remain those of the Lorentzian progenitor \eqref{eqn:explicitmet}. 

The BPS geometries are complex for generic values of the three real parameters $(a, r_+, r_*)$, but most of their building blocks remain real. For example, the shifted coordinates $r_{1,2}$ introduced in \eqref{eqn:shifteddef} are real, and $\Delta_\theta, \rho^2$ given in \eqref{eqn:delthetadef} are real as well. It is only $\Delta_r^*$ \eqref{eqn:DerBPS} that is genuinely complex. Moreover, its imaginary part is proportional to $r_+^2-r^2_*$, so the entire metric is real in the limit
$r_+^2\to r^2_*$. Therefore, the KSW conditions \eqref{eqn:KSWBPS2} reduce to\footnote{The KSW condition \eqref{eqn:KSWmet} for $p=2$ does not give any additional constraints, because the metric is block diagonal. In this case we can restrict the $p=2$ condition to the upper $2\times 2$ block: ${\rm Re}\, [\sqrt{-\det g} \, (g^{tt}g^{\phi\phi}-(g^{t\phi})^2)]<0$. This condition is equivalent to ${\rm Re}\, \beta>0$.} 
\begin{equation}
{\rm Re} \beta > 0~,
\label{eqn:KSWBPS3}
\end{equation}
on the real BPS surface. This is equivalent to the condition \eqref{eqn:Omegacond} which was identified multiple times already, including by simply demanding positive temperature ${\rm Re}T>0$. Interestingly, this does not capture the stronger condition \eqref{eqn:chargeasym} imposed by the sign of the electric potentials 
${\rm Re} \Phi_{1, 2} > 0$. Compared with the region \eqref{eqn:argTrange} for ${\rm Arg}T$ required by a Hilbert space interpretation, the KSW conditions impose the upper limit, but not the lower limit.\footnote{It was previously reported that, for AdS$_5$ black holes with two angular momenta and a single charge, the KSW condition is necessary, but not sufficient \cite{BenettiGenolini:2025jwe}. However, the evidence presented in \cite{BenettiGenolini:2025jwe} indicates that the KSW condition {\it is} sufficient on the real BPS surface (at $r_+=r_*$). We find that, for AdS$_4$ black holes with one angular momentum and two charges, the KSW conditions are insufficient on the real BPS surface.} 

In some examples, it is useful to write the metric in a diagonal basis as $g = \lambda_i \delta_{ij}$ and formulate the KSW conditions \eqref{eqn:KSWmet} as:
\begin{equation}
\sum_k | {\rm Arg} \lambda_k |<\pi~. 
\label{eqn:KSWlambda}
\end{equation}
This criterion must be applied carefully. 
The Euclidean metric is obviously not Hermitean, or else it would not have complex eigenvalues in the first place. Moreover, its real and imaginary parts do not commute, so no orthogonal transformation of the real Euclidean coordinates can diagonalize them simultaneously. These operations are not possible, and they are not called for. 

Because \eqref{eqn:KSWBPS2} is satisfied, the combination of an orthogonal transformation and a real rescaling of the coordinates can take ${\rm Re}~ g_{ij}$ to the identity matrix. Subsequently, we can diagonalize ${\rm Im} ~g_{ij}$. In the resulting basis, the metric takes the diagonal form $g_{ij} = \lambda_i \delta_{ij}$. The $\lambda$'s are not the eigenvalues of the quadratic form $g_{ij}$, because the rescaling of the coordinates leave the ``size" of each direction indeterminate. Indeed, we recognized in advance that conventional diagonalization is impossible. The best we can do is to identify the phase of $\lambda_i$, but not its magnitude. 

 After taking the real part, the matrix in \eqref{eqn:KSWBPS2} is automatically diagonal when the black holes are on the real BPS surface. We then 
 rescale the basis vectors by real factors $t\to  \frac{1}{\sqrt{-g_{tt}{{\rm Re} \beta}}} t$, $\phi\to \frac{1}{\sqrt{g_{\phi\phi}{\rm Re} \beta^{-1}} }\phi$, and similarly for $r$ and $\theta$. These
 transformations give:  
 \begin{equation}
\frac{1}{\sqrt{-g}}g_{ij} ~ \to~     \frac{1}{{\rm Re} \beta} \begin{pmatrix} \beta  & \frac{- i g_{t\phi}|\beta|}{\sqrt{-g_{tt} g_{\phi\phi}}} & 0 & 0  \cr  \frac{- i g_{t\phi}|\beta|}{\sqrt{-g_{tt} g_{\phi\phi}}} & \beta^*& 0 & 0 \cr 0 & 0 & \beta^* & 0\cr 0 & 0 &  0 & \beta^* 
 \end{pmatrix} . 
\label{eqn:KSWBPS4}
\end{equation}
The left hand side is the original Lorentzian metric. The $\to$ indicates the combination of continuation to Euclidean signature $t\to - i \beta t$ and real rescaling of the coordinates. The real part of the right hand side is the identity matrix. The upper left $2\times 2$ of the imaginary part is a symmetric matrix with zero trace and negative determinant. It has eigenvalues $\pm\kappa$ where
\begin{equation}\label{kappa}
    \kappa = \frac{1}{{\rm Re} \beta} \sqrt{ {\rm Im}^2\beta  - \frac{g_{t\phi}^2|\beta|^2}{g_{tt} g_{\phi\phi}}}~.
\end{equation}
After a real orthogonal transformation that diagonalizes the matrix, the four diagonal entries of \eqref{eqn:KSWBPS4} become
$$
\frac{1}{\sqrt{-g}}g_{ij}  \to    \Big\{  1  +  i \kappa~,~1  - i \kappa ~,~
  \frac{\beta^*}{{\rm Re} \beta} ~,~ \frac{\beta^*}{{\rm Re} \beta} \Big\}~.
$$
The continuation to Euclidean signature took $\sqrt{-g}  \to  \beta\sqrt{g}$ so, up to real factors, the final result after all transformations becomes: 
\begin{equation}
g_{ij} \to    \Big\{   \beta \left( 1  + i \kappa \right)~,~
 \beta \left(  1  - i \kappa  \right)~,~
 1 ~,~ 1 \Big\}~.
 \label{eqn:lambdafinal}
\end{equation}
These are the $\lambda$'s that enter the condition \eqref{eqn:KSWlambda}. The KSW condition in terms of the eigenvalues $\kappa$ reads
\begin{equation}\label{kswfinal}
    |{\rm Arg} \, \beta + {\rm Arg} \, (1+i\kappa)|+|{\rm Arg} \, \beta - {\rm Arg} \, (1+i\kappa)| < \pi ~,
\end{equation}
which is equivalent to the conditions
\begin{equation}\label{equivksw}
    |{\rm Arg} \, \beta|<\frac{\pi}{2} \quad \text{ and } \quad |{\rm Arg} \, (1+i\kappa)|<\frac{\pi}{2}~.
\end{equation}
The eigenvalue $\kappa$ is always real, since the expression in the square root of \eqref{kappa} is positive everywhere outside the horizon. Therefore the second condition in \eqref{equivksw} is true everywhere outside the horizon. We therefore conclude again that the only constraint imposed by KSW condition is $|{\rm Arg} \,\beta|<\frac{\pi}{2} $ which is equivalent to ${\rm Re} \, \beta >0$. This is necessary, to be sure, but it is not sufficient.

\section{Conclusion}
\label{sec:conclusion}
The main purpose of this article is to develop the phase diagram of supersymmetric states in AdS$_4$. We work entirely in the large-$N$ approximation where the dominant BPS states involve a black hole. Pure, single component BPS black holes exist only on the black hole sheet, a codimension one surface in the three dimensional space of charges parametrized $(J, Q_1, Q_2)$. For other charges, the prevailing BPS states are two-component configurations comprising a core black hole and a ``gas" which, in the classical approximation, carries macroscopic charge but negligible entropy. 

Our main results are: 
\begin{itemize}
    \item We constructed the phase diagram for the supersymmetric {\it ground states} presented in Figure \ref{fig:cohom}. The green line represents the black hole sheet. Region I is ``over-rotating", and the gas carries angular momentum while Region II is ``over-charged" and the gas carries electric charge. The latter is divided into Region $II_a$ where the core black hole has $Q_1=Q_2$ and Region $II_b$ where $Q_1\neq Q_2$. 
    \item We constructed the phase-diagram for the {\it index} presented in Figure \ref{fig:indphasediag}. Any index line that crosses the physical region intersects the black hole sheet at precisely one point, so there is always a specific pure BPS black hole that is a candidate for domination of the index. Our phase diagram shows that, when the two R-charges are sufficiently unbalanced, the index is dominated by a two-component configuration. 
    \item We identified the complex BPS black holes in some detail. The black hole sheet is obtained in the limit where geometries and charges become real, but potentials remain complex. We analyzed the KSW criterion for the Euclidean analytic continuation of the complex BPS black holes and found that it does not capture the predicted instability in the saddle point of the Index. 
\end{itemize}

As is often the case, many interesting questions are left for future research. Some of them are: 
\begin{itemize}
\item 
We did not discuss the {\it nature of the ``gas" component} in any detail, we only made general remarks in subsection \ref{subsec:gas}. It is important to show that these configurations exist, in both region I and region II, and that they have the properties we assume, including supersymmetry. 
\item 
We did not analyze {\it the index of the dual CFT$_3$}. It would be interesting to identify the phase transitions we predict, at the lines where we expect them. 
\item 
The phase transitions we identify are of second order, so they should be subject to large critical fluctuations and described by a universal {\it effective quantum field theory} that can be identified from the symmetries of the problem. 
\item The significance of {\it complex BPS solutions} remains unclear. The complex family we construct amounts to a generalization of the known real BPS black holes that depends on one more real parameter. The two component configurations that we propose as BPS ground states also depend on one more real parameter, but they have no obvious relation to the complex solutions. 
\item 
It would be desirable if the criteria for {\it allowable Euclidean solutions} excluded precisely those pure BPS black holes that fail to dominate the index line they belong to. The KSW conditions do not have this property, but it may be realized by a variation over the same idea. 
\item Many {\it generalizations} of our work are possible. For example, it is worthwhile to study the phase diagram of non-extreme black holes, four independent electric charges, quantum corrections in the AdS$_2$ throat, and higher derivative corrections. 
\end{itemize}
It would be interesting to pursue these research directions. 

\section*{Acknowledgements}

We thank Areliz Tamayo for collaboration in the initial stages of this project. We also thank
L. Pando Zayas,
C. Patel, 
and 
K. Sharma
for discussions.
This work was supported in part by DoE grant DE-SC0007859, the Leinweber Institute for Theoretical Physics, and the Department of Physics at the University of Michigan. 

\appendix

\section{Entropy function extremization for pure BPS black holes}\label{sec:entext}

In this appendix we compute the indicial entropy, by extremization of the indicial free energy followed by Legendre transformation. 
By solving the constraint from the outset, it will be menifest that our computation depends only on the indicial variables $(\zeta, \tilde\zeta)$.

The starting point for the standard derivation of the 
indicial entropy $S$ is the indicial free energy \cite{Choi:2018fdc, Larsen:2020lhg}
\begin{equation}\label{freeindapp}
    \mathcal F = -ik \frac{\Phi_1^\prime \Phi_2^\prime}{\Omega^\prime}~,
\end{equation}
that is subject to the constraint: 
\begin{equation}\label{indconstapp}
    \Phi_1^\prime + \Phi_2^\prime -  \Omega^\prime = 2\pi i~.
\end{equation}
The entropy follows by performing the extremization
\begin{equation}\label{extapp}
    S_{\mathcal{I}} = {\rm ext}_{\Phi_1^\prime,\Phi_2^\prime,\Omega^\prime,\Lambda } \left\{-\mathcal F  + k ( 2\Phi_1^\prime q_{1} + 2\Phi_2^\prime q_{2} + \Omega^\prime j) + \Lambda( \Phi_1^\prime + \Phi_2^\prime - \Omega^\prime - 2\pi i)\right\}~.
\end{equation}
where the constraint \eqref{indconstapp} is implemented by a Lagrange multiplier $\Lambda$. Assuming real charges $(q_{1},q_{2},j)$ and real entropy, the result
\begin{equation}
    S_{BH}(q_{1c},q_{2c}) = k\pi \left(\sqrt{1+16 q_{1c}q_{2c}}-1\right)~,
\end{equation}
agrees with the entropy of a pure BPS black holes when 
its charges $(q_{1c},q_{2c},j_c)$ is identified with the charges $(q_1,q_2,j)$ that appear in \eqref{extapp}.
This agreement between $S$ and $S_{BH}$ is modulo the charge constraint 
\begin{equation}
\label{eqn:jcapp}
    j_c = (q_{1c}+q_{2c})\left(\sqrt{1+16 q_{1c} q_{2c}} -1\right)~,
\end{equation}
that is satisfied for all pure BPS black holes.
The extremization procedure also computes the complex chemical potentials $\Phi_1^\prime$, $\Phi_2^\prime$, and $\Omega^\prime$ as function of the charges $(q_{1c}, q_{2c})$ of the pure BPS black hole \cite{Larsen:2020lhg} 
\begin{align}
    \frac{1}{2\pi}{\rm Re }~ \Phi_{1,2}^\prime &=  \frac{(q_{1c}+q_{2c})\mp (q_{1c}-q_{2c})\sqrt{1+16q_{1c}q_{2c}}}{1+4\left[(q_{1c}+q_{2c})^2+4q_{1c}q_{2c}\right]} ~, \label{chempotph12re}\\
    \frac{1}{\pi}{\rm Im }~ \Phi_{1,2}^\prime &= -1+ \frac{\pm 4(q_{1c}^2-q_{2c}^2)+\sqrt{1+16 q_{1c}q_{2c}}}{1+4\left[(q_{1c}+q_{2c})^2+4q_{1c}q_{2c}\right]} ~,\label{chempotph12im}\\
    \frac{1}{4\pi}{\rm Re }~ \Omega^\prime &=  \frac{ q_{1c}+q_{2c}}{1+4\left[(q_{1c}+q_{2c})^2+4q_{1c}q_{2c}\right]}~,\label{chempotomre}\\
    \frac{1}{4\pi}{\rm Im }~ \Omega^\prime &= \frac{\sqrt{1+16 q_{1c}q_{2c}}}{1+4\left[(q_{1c}+q_{2c})^2+4q_{1c}q_{2c}\right]} ~. \label{chempotomim}
\end{align}

The extremization principle \eqref{extapp} is expressed in terms of charges that satisfy the nonlinear constraint \eqref{eqn:jcapp},
rather than the indicial charges $\zeta = q_1+q_2+j$ and $\tilde\zeta = q_1-q_2$. To address this shortcoming, we perform the extremization over $\Lambda$ first or, equivalently, solve the 
constraint \eqref{indconstapp} from the outset. This gives an 
extremization principle 
\begin{align}
    S_{\mathcal{I}} 
    &= {\rm ext}_{\Phi_1^\prime,\Phi_2^\prime} \left\{-\mathcal F + k ~\Phi_1^\prime (\zeta+\tilde\zeta) + k ~\Phi_2^\prime (\zeta-\tilde\zeta) \right\}~,\label{extmanapp}
\end{align}
that manifestly depends on the real indicial charges $(\zeta,\tilde\zeta)$ only. In the second line we omitted the purely imaginary term $-2\pi i kj$ that does not affect the real part of the entropy $S$. In \eqref{extmanapp} the function $\mathcal F$ 
\begin{equation}
    \mathcal{F} = -ik \frac{\Phi_1^\prime \Phi_2^\prime}{\Phi_1^\prime + \Phi_2^\prime - 2\pi i}~,
\end{equation}
that follows is by inserting the constraint \eqref{indconstapp} in \eqref{freeindapp}.

The saddle point equations for the  \eqref{extmanapp} are
\begin{align}
    \frac{\partial \mathcal F}{\partial \Phi_1^\prime} = -ik \frac{\Phi_2^\prime (\Phi_2^\prime - 2\pi i)}{(\Phi_1^\prime + \Phi_2^\prime - 2\pi i)^2} = k (\zeta+\tilde\zeta)~,\\
    \frac{\partial \mathcal F}{\partial \Phi_2^\prime} = -ik \frac{\Phi_1^\prime (\Phi_1^\prime - 2\pi i)}{(\Phi_1^\prime + \Phi_2^\prime - 2\pi i)^2} = k (\zeta-\tilde\zeta)~.
\end{align}
These equations can be rewritten as
\begin{align}
    \zeta &= -\frac{i}{2} \left[\frac{\Phi_1^\prime+\Phi_2^\prime}{\Phi_1^\prime + \Phi_2^\prime - 2\pi i}- \frac{2\Phi_1^\prime \Phi_2^\prime}{(\Phi_1^\prime + \Phi_2^\prime - 2\pi i)^2}\right] \label{zeta1eq}~,\\
    \tilde\zeta &= -\frac i2 \left[\frac{\Phi_2^\prime - \Phi_1^\prime}{\Phi_1^\prime + \Phi_2^\prime - 2\pi i}\right]~.
\end{align}
The second equation gives $\Phi_2^\prime$ as a linear function of $\Phi_1^\prime$
\begin{align}
    \Phi_2^\prime = \Phi_1^\prime \left(\frac{i-2\tilde\zeta}{i+2\tilde\zeta}\right) + 2\pi i \left(\frac{2\tilde\zeta}{i+2\tilde\zeta}\right)~,
\end{align}
which, when substituted in \eqref{zeta1eq}, gives a quadratic equation for $\Phi_1^\prime$. Solving for the chemical potentials $\Phi_{1,2}^\prime$ we find
\begin{align}
    \Phi_1^\prime &= i\pi \left(-1 + \frac{1+2i\tilde\zeta}{\sqrt{1-4\tilde\zeta^2+4i\zeta}}\right)~,\label{appphi1p} \\
    \Phi_2^\prime &= i\pi \left(-1 + \frac{1-2i\tilde\zeta}{\sqrt{1-4\tilde\zeta^2+4i\zeta}}\right)~. \label{appphi2p}
\end{align}
Among two possible solutions, we picked the branch so the entropy computed below is positive. As a consistency check on our computations, we verified that (\ref{appphi1p}-\ref{appphi2p}) agree with the chemical potentials (\ref{chempotph12re}-\ref{chempotph12im}) written as functions of BPS black holes charges, upon relating
the charges of the BPS black hole at point $X$ and the indicial charges $(\zeta,\tilde\zeta)$ through (\ref{qxcoord}-\ref{qtildexcoord}).  

 Substituting the results (\ref{appphi1p}-\ref{appphi2p}) for the chemical potentials into the exttremization principle \eqref{extmanapp}, and extracting the real part of the result, we get
\begin{align}
    S_{\mathcal{I}}(\zeta,\tilde\zeta) = k\pi \left(-1 +  \sqrt{\frac12\left(1 - 4 \tilde\zeta^2+ \sqrt{\left(1 - 4 \tilde\zeta^2\right)^2+16 \zeta^2}\right)} \right)~. 
\end{align}
This formula agrees with the entropy \eqref{indxent} of the pure BPS black hole at the interesection of the black hole sheet and the indicial line.

\bibliographystyle{JHEP}
\bibliography{refs.bib}

\end{document}